\documentclass[12pt]{iopart}
\usepackage{graphicx}

\begin{document}
\title[Moment analysis for QES]{A Moments'   Analysis  of Quasi-Exactly Solvable Systems: A New Perspective on the Sextic Anharmonic and  Bender-Dunne Potentials}
\author{Carlos  R. Handy$^1$, Daniel Vrinceanu$^1$, and Rahul Gupta$^2$}
\address{$^1$Department of Physics, Texas Southern University, Houston, Texas 77004; \\$^2$Lawrence E. Elkins High School, Missouri City, Texas 77459}
\ead{handycr@tsu.edu}

\begin{abstract}
There continues to be great interest in understanding quasi-exactly solvable (QES) systems. In one dimension, QES states assume the form $\Psi(x) =x^\gamma P_d(x) {\cal A}(x)$, where ${\cal A}(x) > 0$ is known in closed form, and $P_d(x)$ is a polynomial to be determined. That is ${{\Psi(x)}\over {x^\gamma{\cal A}(x)}} = \sum_{n=0}^\infty a_nx^n$ truncates. The extension of this ``truncation" procedure to non-QES states corresponds to the Hill determinant method, which is unstable when the {\it reference} function assumes the physical asymptotic form (i.e. $x^\gamma{\cal A}(x)$). Recently, Handy and Vrinceanu introduced the Orthogonal Polynomial Projection Quantization (OPPQ) method which has non of these problems, allowing for a unified analysis of QES and non-QES states ( 2013 J. Phys. A: Math. Theor. {\bf{46}} 135202; 2013 J. Phys. B: {\bf{46}} 115002). OPPQ uses  a non-orthogonal basis constructed from the orthonormal polynomials of ${\cal A}$: $\Psi(x) = \sum_{j=0}^\infty \Omega_j {\cal P}^{(j)}(x) {\cal A}(x)$, where $\langle {\cal P}^{(j_1)}|{\cal A}|{\cal P}^{(j_2)} \rangle = \delta_{j_1,j_2}$, and $\Omega_j = \langle {\cal P}^{(j)}|\Psi\rangle$. For systems admitting a moment equation representation, such as those considered here, these coefficients can be readily determined. 
The OPPQ quantization condition, $\Omega_{j} = 0$, is exact for QES states (provided $j \geq d+1$); and is computationally stable, and exponentially convergent, for non-QES states. OPPQ provides an alternate explanation to the Bender-Dunne (BD) orthogonal polynomial formalism for identifying QES states: they correlate with an anomalous kink behavior  in the order of the finite difference moment equation associated with the $\Phi = x^\gamma {\cal A}(x) \Psi(x)$ {\it Bessis}-representation (i.e. a spontaneous change in the degrees of freedom of the system). This was first noted by Handy and Bessis in their  implementation of the Eigenvalue Moment Method (EMM), the first application of semidefinite programming analysis to quantum operators (1985 {\em Phys. Rev. Lett.} {\bf 55}, 931 ).  Additional properties ensue, such as $\Phi_{non-QES}(x) = \partial_x^{d+2}\Upsilon(x)$, for states of the same symmetry as the QES states. We study the above with respects to two sextic potentials of the type $V(x) = gx^6+bx^4+mx^2 +{\beta\over {x^2}}$.
\end{abstract}
\submitto{\JPA}

\vfil\break
\section{Introduction}

\subsection {Objectives and Overview}

\bigskip\noindent\underbar {\it{Inadequacies of the Hill determinant representation}}
\bigskip

The study of quasi-exactly solvable (QES) systems has continued to attract much interest because of its relevance to physical systems and extensions to quantum supersymmetry [1]. These correspond to  Hamiltonians  for which a subset of the discrete spectrum, and corresponding wavefunctions, can be determined in closed form. Their systematic study was initiated by Turbiner [2-5].  In one space dimension, the typical QES state corresponds to a wavefunction of the form $\Psi(x) = P_d(x) {\cal A}(x)$, where the positive asymptotic configuration ${\cal A}(x)> 0 $ is known in closed form, and $P_d(x)$ is some polynomial of degree `$d$' , to be determined. In other words, the ratio ${{\Psi(x)}\over {\cal A}(x)}$ truncates. This truncation philosophy does not naturally extend, as given, to the non-QES states, for reasons given below. This is the primary objective of this work:  to develop a unified theoretical and computational framework that can address the QES and non-QES states. Additionally, our methods give a different interpretation  for the existence of QES states, from a moments' representation perspective.

One may regard the QES truncation philosophy as a motivating factor for the general Hill determinant quantization philosophy [6] which, in one dimension, represents an arbitrary discrete state as  $\Psi(x) = x^\gamma A(x) {\cal R}(x)$, where $\gamma$ is the (problem dependent)  indicial exponent, $A(x) = \sum_{j=0}^\infty a_j x^j$ is an analytic factor, and ${\cal R}(x)$ is some specified {\it{reference}} function, such as the Gaussian, $e^{-x^2}$. One can relate the analytic properties of the wavefunction to the $a_j$'s, which also acquire an energy dependence. The Hill determinant quantization prescription determines those (approximate) energies  leading to an effective truncation of the power series expansion, $a_N(E) = 0$, etc. In the limit $N\rightarrow \infty$, these energy approximants (for the non-QES states) usually converge to the true physical values, for the Gaussian reference function. The major drawback of this approach, as is well known, is that  if the reference function is chosen to mimic the true asymptotic form of the wavefunction, it leads to instabilities and erroneous energy convergence [7]. For the sextic anharmonic oscillator potential, $V_{sa}(x) = gx^6+ b x^4 + m x^2$,  and  the Bender Dunne [8] sextic potential, $V_{BD}(x) =x^6 +mx^2 + {b\over {x^2}}$,  the physical  reference functions are ${\cal R}_{sa}(x) = e^{-{{\sqrt {g}}\over 4} \big(x^4 +{b\over g} x^2\big)}$ and ${\cal R}_{BD}(x) = e^{-{{x^4}\over 4}}$,  respectively. However, the Hill determinant approach proves unstable in either case, as suggested by the  study by Tate and Turbiner [9].  Thus,  both QES and non-QES states can be approximated, if a Gaussian type reference function is used. If the true asymptotic form for the physical states could be used as reference functions, then the same quantization approach would generate both the exact QES states and approximate the non-QES states. However, this is inherently impossible within the Hill determinant  ``truncation" philosophy. Nevertheless, the methods introduced here can do precisely this.

Of relevance to the present formalism is the fact that the Hill determinant method can also be implemented in Fourier space, ${\hat{\Psi}}(k) = k^\gamma A(k) {\cal R}(k)$. It is referred to as the Multiscale Reference Function (MRF) method, originally proposed by Tymczak, Japaridze, Handy, and Wang  [10]. The selection of an appropriate, positive, reference function is somewhat limited, with the only viable choice usually being the Gaussian, ${\cal R}(k) = e^{-k^2}$. However, even for this case, the MRF method has better (faster and more monotonic) convergence properties than the  configuration space Hill determinant approach. A comparison, for the sextic anharmonic oscillator, is given in Ref. [11].  The MRF method is only implementable if the Schrodinger equation admits a moment equation representation. If so, then the power moments of the (discrete states), $\mu(p) = \int dx \ x^p \Psi(x)$, satisfy a linear, recursive relation that involves the energy as a variable parameter. These are then used to generate the power series coefficients of $A(k)$. The relevance of the MRF method to the present analysis is that what is proposed here can be considered as a merging of the Hill and the MRF into a new and much more powerful representation particularly relevant for QES systems, as well as exactly solvable systems (i.e. for which all discrete states are determinable in closed form), which will be discussed in a subsequent work. 

\bigskip\noindent\underbar {\it{ Orthogonal Polynomial Projection Quantization}}
\bigskip

Recently, Handy and Vrinceanu [11] proposed a new, multidimensional, quantization formalism for systems admitting a moment equation representation. It is referred to as the Orthogonal Polynomial Projection Quantization (OPPQ) method. A motivatng factor  was simply to improve upon the known limitations of the Hill determinant analysis. These limitations not only include the aforementioned  instabilities when the reference function mimics the physical asymptotic configuration (${\cal R}(x) \rightarrow {\cal A}(x)$), but also the requirement that the reference function be analytic (of importance to the Bender Dunne potential). Neither of these is a limitation within OPPQ. 

The implementation of OPPQ  requires working within a non-orthogonal  basis, $\{{\cal P}^{(j)}(x){\cal R}(x)| j \geq 0 \}$,  formed from the orthonormal polynomials of the positive reference function, ${\cal R}(x)$: $\langle {\cal P}^{(j_1)}|{\cal R}|{\cal P}^{(j_2)}\rangle = \delta_{j_1,j_2}$. These are used to generate the  representation: $\Psi(x) = \sum_{j=0}^\infty \Omega_j {\cal P}^{(j)}(x){\cal R}(x)$. The expansion coefficients project out exactly, $\Omega_j = \langle {\cal P}^{(j)}|\Psi\rangle$. They correspond to finite sums of the power moments of $\Psi$. One can easily argue that for a broad class of reference functions, including those asymptotic to the discrete physical states, we must have $\lim_{n\rightarrow \infty}\Omega_n = 0$. Assuming the existence of a moment equation (of effective order $1+m_s$), the $\Omega_j$'s become linearly dependent on the first $1+m_s$ power moments through known, energy dependent, coefficients. Therefore, one can define the OPPQ quantization procedure as taking $\Omega_{N+\ell} = 0$, for $\ell = 0,\ldots,m_s$. This yields an energy dependent determinantal equation, $D_N(E) = 0$, whose roots exponentially converge to the physical states in the  $N\rightarrow \infty$ limit.

 If the  asymptotic configuration of the physical states is known in closed form, $x^\gamma{\cal A}(x) > 0$ (i.e. if $\gamma \neq integer$, then $x\rightarrow 0^+$ and $x\rightarrow \infty$ are necessary asymptotic  limits), then one can take it to be the reference function: ${\cal R}(x) = x^\gamma{\cal A}(x)$. In this case, a subset of the OPPQ determinant's roots  are the exact QES energies, for $N$ greater than a certain threshold.  Specifically, for  one dimensional systems, since a QES state must assume the form $\Psi_{QES}(x) = P_d(x) x^\gamma{\cal A}(x)$, and
$P_d(x) = \sum_{j=0}^d c_j {\cal P}^{(j)}(x)$, then $\Omega_n = \langle {\cal P}^{(n)}|\Psi_{QES}\rangle = 0$, for $n \geq d+1$.  The QES energies must be the exact roots, $D_N(E_{QES}) = 0$ for  $N \geq d+1$, where each QES state will have its own ``$d$". The non-QES states are approximated by the other roots of the OPPQ determinant; and these approximations converge exponentially fast to the physical values. 

Although this work is limited to one dimensional systems, the entire OPPQ formalism can be, and has been, applied in two dimensions. The original work by Handy and Vrinceanu investigated several two dimensional quantum systems including a particular pseudo-hermitian model. A subsequent work applied the OPPQ formalism to the challenging 
two dimensional  infinite  dipole problem  (two oppositely charged line charges) [12] confirming a large basis Rayleigh Ritz analysis by Amore and  Fernandez  [13].

In summary, the Hill determinant truncation philosophy (trivially) works for QES states, but is unstable, or ineffective for the non-QES states. The OPPQ analysis was developed independently of QES considerations, but turns out to be the ideal, unifying quantization framework for both QES and non-QES, as presented here.
\vfil\break
\bigskip\noindent\underbar{{Obtaining the QES (Bender Dunne), Energy, Polynomials}}
\smallskip

Within a configuration space Hill representation, ${{\Psi(x)}\over {x^\gamma {\cal A}(x)}} = \sum_n\ a_n(E)x^n$, if the potential function parameters satisfy a particular constraint, then the $a_n$'s will exhibit the defining truncation structure, $a_n(E) = 0$, for $n\geq n_*+1$, where $d \equiv n_*$;  and for which  $a_{n_*+1}(E) $  is a polynomial of degree $n_*+1$ whose roots correspond to all the QES states for that system. This polynomial is contained in the OPPQ determinant expression; although its forms is not as immediately discernable as it is in the Hill representation for the $a_n$'s.  Whereas the Hill representation is inadequate for determining the non-QES states, the OPPQ-$\Psi$ analysis is able to generate both QES and non-QES states in a unified manner. 

We would like a moments' representation where the $a_{n_*+1}(E)$ polynomial is also immediately transparent, and both QES and non-QES states can be generated in a unified manner through OPPQ.  This is possible within the {\it{Bessis}} representation defined by $\Phi(x) = x^\gamma {\cal A}(x) \Psi(x)$. The OPPQ-$\Phi$ analysis now requires working with reference functions of the type ${\cal R}(x) = x^{2\gamma}{\cal A}^2(x)$, and their orthonormal polynomials.  Note that contratry to the Hill representation, we are not stripping the asymptotic form(s), but further enhancing the new representation by these expressions. Within this representation, the corresponding moments, $\nu(p) = \int dx \ x^p\Phi(x)$, have a moment equation that: (i) makes the identification of the $a_{n_*+1}(E)$ polynomial very transparent; and (ii) exhibits a structure that undelies the real reason for the  anomalous behavior of the Bender-Dunne (energy dependent) orthogonal polynomials. 

Within the  Hill representation, the expansion coefficients, $a_n(E)$, satisfy a three term recursion relation,  becoming polynomials in the energy variable. This recursion relation remains of first order (i.e. $a_n$ generates $a_{n+1}$, etc.)  for all `$n$', regardless of the energy (QES or non-QES). One can transform the $a_n$-recursion relation so that it resembles the monic form of the usual three term relation for orthogonal polynomials.   Thus, $a_n(E)$, an $n$-th degree polynomial, transforms into an $n$-th degree polynomial, $P_{bd}^{(n)}(E) \propto a_n(E)$, which satisfies the manifestly monic three term relation $P_{bd}^{(n)}(E) = (E-\alpha_n) P_{bd}^{(n-1)}(E) - \gamma_{n-1} P_{bd}^{(n-2)}(E)$. What Bender and Dunne did was to reinterpret the existence of QES states, within the $P_{bd}$-representation, as corresponding to a breakdown of this recursion relation, with $\gamma_{n_*+1} =0$. This relation implies that these monic orthogonal polynomials have a signed weight, $w(E)$, and relative to that weight the quantizing polynomial has zero norm: $\langle P_{bd}^{(n_*+1)}|w|P_{bd}^{(n_*+1)}\rangle = 0$.

From the moments' perspective, the Bender and Dunne interpretation is not necessary because the form of the moment equation for the $\nu$'s reveals the true anomaly.

 For the non-QES states of different symmetry to the QES states (i.e. the sextic anharmonic oscillator case) ,  the $\nu$-moment equation is also a three term, first order, recursion relation. However, for the QES states, the moment equation is only of first order for the first $n_*+1$ moments, $\{\nu(p)|0 \leq p \leq n_*\}$.  All the other moments satisfy a finite difference equation of second order.  Within the context of the moment equation, the $\nu(n_*+1)$ moment decouples from the lower order moments; although, it can be generated from the lower order moments through a relation independent of the moment equation. For the QES states within the same symmetry class as the QES states, the disruption is less severe since the first $n_*+1$ moments must be zero (i.e. $\nu(p) = 0,\  \leq p \leq n_*$); whereas all the higher order moments define a first order finite difference equation. This {\it{kink}} in the order of the underlying moment equation for QES states is the real reason for the Bender-Dunne orthogonal polynomial anomaly. 

For the QES states, the $\{ \nu(p)|0 \leq p \leq n_*\}$ moments are polynomials in the energy, $E$, of degree corresponding to the moment-order (`$p$').  The BD polynomial $P_{bd}^{(n+1)}(E)$ corresponds to the linear sum of the two moments $\{\nu(n-1),\nu(n)\}$, involving an energy dependent coefficient that is a monomial in the energy (consistent with the monic form for orthogonal polynomials). Because of the kink in the $\nu$-moment equation, from the moments' perspective, the $P_{bd}^{(n+1)}(E)$ polynomials have no natural extension beyond $n \geq n_*$. For the non-QES states, for symmetry different than the QES states, the corresponding monic orthogonal polynomials exist (i.e. satisfy the monic orthogonal polynomial three term structure). 

In summary, we do not have to look for kinks in the recursive structure of the Bender Dunner orthogonal polynomial representation, it is easier to look for kinks in the nature of the moment equation within the Bessis representation. The latter would  seem to be an easier analysis than the former, particularly for multidimensional systems admiting a moment equation representation.

\bigskip\noindent\underline{\it{The Bessis Representation: Relevance of the Eigenvalue Moment Method}}
\bigskip

The existence of QES states as due to  a nonuniform moment order structure was known to Handy and Bessis (HB)  in the context of their development of the Eigenvalue Moment Method (EMM) [14-16], the first application of semidefinite programming (SDP) [17-18] in quantum physics. We ouline the relevant history since it impacts this work, and underscores the importance of moment representations for quantizing physical systems. 

In 1984 Handy [19] discovered that combining the moment equation representation with a particular representation of the {\it{ moment problem}} theorems in mathematics  (i.e. the nesting property of the Pade approximants of Stieltjes measures) [20], yielded converging lower and upper bounds to the (one dimensional) bosonic ground state energy, or any other quantum state associated with a  known nodal structure  (i.e. the first excited state of parity invariant systems). Handy and Bessis (HB)  transformed this into a more general, multidimensional, formulation through the use of the moment problem  Hankel Hadamard (HH) determinantal  inequality constraints, $Det({\cal H}_n) > 0$, where ${\cal H}_{i,j} = \mu(i+j), 0 \leq i,j \leq n$, is the Hankel matrix [21]. Through the underlying moment  equation of order $1+m_s$, these moments depend linearly on the $1+m_s$ initialization moments (i.e. referred to as the {\it{missing moments}} by HB), and nonlinearly on the energy parameter.   Unlike typical SDP problems that seek to optimize some {\it{objective}} function, the EMM/SDP analysis requires that for each energy parameter value, $E$, one  determine the existence or nonexistence of the nonlinear convex solution set to the HH inequalities (once a suitable normalization condition is chosen, reducing the missing moment domain to  $m_s$ dimensions), $Det\Big({\cal H}_n({\cal U}_E^{(N)})\Big) > 0$, $n \leq N$, $N \rightarrow \infty$.   The set of admissible energy values define an interval $[E_L^{(N)},E_U^{(N)}]$, such that if $E_L^{(N)} \leq  E \leq E_U^{(N)}$, then  ${\cal U}_E^{(N)} \neq \emptyset$. As the dimension of the Hankel matrix increases, $N \rightarrow \infty$, we have $E_U^{(N)}-E_L^{(N)} \rightarrow 0$, exponentially with `$N$', in most cases. This bounding procedure was particularly effective for strongly coupled, singular perturbation type systems for which conventional computational methods could be unreliable.  

The EMM analysis, as originally formulated [14], is a nonlinear optimization problem. There were no efficient SDP algorithms available in 1985, thus limiting the class of problems HB  could investigate. The hardest problem amenable to a very basic computational strategy was the sextic anharmonic oscillator.  Motivated by this, Bessis realized that by multiplying the wavefunction by its asymptotic form  the computational complexity of the sextic anharmonic oscillator problem became equivalent to that for the harmonic oscillator problem. This revealed the anomalous kink behavior of the $\nu$-moment equation for the QES states; however the focus of that first work was on bounding the non-QES energies for the ground and first excited states.   In subsequent works  [15,16] Handy was able to transform the nonlinear version of EMM into an equivalent linear programming based formulation which allowed for its implementation to a broad range of multidimensional systems, including the notoriously difficult quadratic Zeeman effect for superstrong magnetic fields [15,16]. The relevance of moment representations for quantizing singular perturbation-strongly coupled systems had been previously noted by Handy [22], in the context of finding a more rigorous alternative  to lattice high temperature expansions in field theory. This early work introduced the scaling transform, whose perturbative structure in the inverse (lattice) scale depends on the power moments. The relevance of this formalism to incorporating wavelets into quantum mechanics was demonstrated in a subsequent work [23]. Coincidentally, one may characterize EMM as an affine map invariant variational procedure, since it optimizes within the affine map invariant space of polynomials. Indeed, the EMM bounds to the quadratic Zeeman effect were highly correlated with the ground state binding energy estimates of an order dependent, conformal, analysis by LeGuillou and Zinn-Justin [24]; and similarly for the three dimensional quantum dot [25]. Beyond EMM, advances in SDP code development have progressed rapidly since the 1990s with the impetus coming from combinatorics [26] and reduced density matrix theory in quantum chemistry [27-28].

A major focus of  EMM analysis is to define new nonnegativity representations for quantum systems. For arbitrary one dimensional systems with real potentials, the probability density satisfies a third order, linear, differential equation, enabling application of EMM bounding methods for all states, depending on the nature of the potential [29]. That is, knowledge of the nodes is not required. The same is true for complex, one dimensional, potentials, because the Herglotz analytic continuation of $|\Psi(x)|^2$ satisfies a fourth order linear differential equation. One can then use EMM to bound the complex quantum parameters. This was used to support the conjecture on the  reality of the discrete spectrum   for the pseudo hermitian potential $V(x) = ix^3$ (i.e. the {\it{Bessis}} conjecture) [30], as previously suggested by Bender and Boetcher [31], and subsequently proved by Dorey et al [32]. It was also used to  computationally predict the correct onset of PT-symmetry breaking for the pseudo-hermitian potential,
$V(x) = ix^3+ iax$ [33].  Other applications include bounding Regge pole parameters relevant to atomic scattering [34,35].

\vfil\break
 
\bigskip\noindent\underbar{\it {Summary of problems and representations to be analyzed}}
\smallskip

The following sections will examine the sextic anharmonic oscillator potential, $V_{sa}(x) = gx^6+bx^4+mx^2$, and the BD sextic potential, $V_{BD}(x) = x^6 + mx^2+{b\over {x^2}}$,  both within the $\Psi$ representation and the {\it Bessis} $\Phi$ representation. In each we show how OPPQ yields both the exact QES states and converging approximants to  the non-QES states. We also show how the $\nu$-moment equation within the {\it Bessis} representation recovers the BD-polynomials and their recursion relation.  It is to be re-emphasized that within the Bessis representation, the QES states can be generated two ways:(1) as the  roots of an energy polynomial defined by the $\nu$-moments (i.e. effectively the BD energy polynomials); (2) as the exact roots of the OPPQ quantization determinant,  for which the remaining roots approximate the non-QES states.

 The following discussion pertains to both potentials, but we limit our remarks to the sextic anharmonic oscillator case. The sextic anharmonic oscillator problem admits non-QES states and QES states, for particular potential function parameter values. Within each parity class, there will be QES states. There will be non-QES states of the same parity as the QES states; whereas all states of the opposite parity will be  non-QES states. These distinctions may complicate our notation. These distinctions do not arise in the BD sextic potential case because it is defined on the nonnegative real axis. 

The sextic anharmonic oscillator Schrodinger equation, in the $\Psi$ representation is

\begin{equation}
-\partial_x^2\Psi(x) + (g x^6 + b x^4 +  m x^2) \Psi(x) = E \Psi(x).
\end{equation}
It admits even and odd parity states as indexed by $\sigma = 0,1$, respectively. The QES states are represented by $\Psi(x) = P_d(x){\cal A}(x)$, for $d \equiv n_* \geq 0$. Their corresponding parity will be denoted by $\sigma_* = 0\ or \ 1$. The  asterisk notation is exclusively used to identify the QES states and relations.
The potential function parameter constraint admitting QES states corresponds to

\begin{equation}
g^{3\over 2}(16n_* + 12 + 8 \sigma_*)+4mg-b^2 = 0.
\end{equation}
This condition can only be explicitly derived within two representations; either from the Hill representation truncation analysis, or the $\nu$-moment equation relation within the Bessis function representation. It can be surmized within the $\Psi$-moment equation representation from a JWKB analysis and then tested through OPPQ. All of these are implemented in the following sections.

We will not give explicit algebraic forms for the QES energies, etc., because the relations to be given are transparent and readily implementable by the interested reader. Instead, we focus on the numerical consistency of our results with the underlying OPPQ theory. We do not focus on wavefunction reconstruction because this is also 
straightforward. Finally, we have streamlined the OPPQ formalism from that originally presented by Handy and Vrinceanu [11-12]. The present formalism emphasizes the orthonormal  polynomials of the physical reference function.

There are several types of polynomials considered here. First is the polynomial factor, $P_d(x)$, defining the QES state. Second are the orthonormal polynomials, ${\cal P}^{(j)}(x)$ corresponding to the positive reference function, ${\cal A}(x)  > 0$. Third are the BD polynomials in the energy space. Fouth will be  the $\nu$-moments which correspond to polynomials in the energy (and whose superposition defines the BD polynomials). The monic form of the OPPQ orthonormal polynomials, ${\cal P}^{(j)}(x)$, will be denoted by ${\tilde{\cal P}}^{(j)}(x)$.

\section {Preliminaries}
\subsection{The $\Psi$-moment equation}

Before continuing with the OPPQ generalities, we note that Eq.(1) can be transformed into a moment equation for the discrete states. Define $\mu(p) \equiv \int dx x^p\Psi(x)$ where we assume $\Psi(x)$ to be implicitly a bound state, asymptotically vanishing at infinity. Multiplying both sides of Eq.(1) by $x^p$ and performing the necessary integration by parts gives the moment equation:

\begin{eqnarray}
g\mu(p+6) = -b\mu(p+4) -m\mu(p+2) + E\mu(p) + p(p-1)\mu(p-2),
\end{eqnarray}
$p \geq 0$.  The even and odd states separate, $\Psi_\sigma(x)$, $\sigma = 0,1$, respectively.  Although not necessary, we prefer to explicitly work within the even and odd representations, in order to reduce the dimensionality of the OPPQ determinant matrix. We denote the power moments for the even or odd states by 
\begin{eqnarray}
u_{\sigma}(\rho) & = \mu(2\rho+\sigma) = \int_{-\infty}^{+\infty} dx x^{2\rho+\sigma} \Psi_\sigma(x) \nonumber \\
& = \int_0^\infty d\xi \xi^\rho \psi_\sigma(\xi), \ where \ \psi_\sigma(\xi) =  {(\sqrt{\xi})^{\sigma-1}} \Psi_\sigma(\sqrt{\xi}) \ ,
\end{eqnarray}
and  $\xi \equiv x^2$,  $\sigma = 0 \ or \ 1$. We note that for the ground and first excited state the $\psi_\sigma(\xi)$ configuration is nonnegative.  The corresponding $u_\sigma(\rho)$-moment equation becomes

\begin{eqnarray}
gu_\sigma(\rho+3) = -bu_\sigma(\rho+2) -m u_\sigma(\rho+1) + Eu_\sigma(\rho) + 2\rho(2(\rho+\sigma)-1))u_\sigma(\rho-1).\nonumber \\
\end{eqnarray}
The effective order of this homogeneous, linear, moment equation is $1+m_s$ where $m_s = 2$, since the moments $\{u_\sigma(0),u_\sigma(1),u_\sigma(2)\}$   must be specified, in addition to the energy, before all the other moments can be generated. Imposing an $L^1$ normalizaiton condition (i.e. $\sum_{\ell = 0}^{m_s} u_\sigma(\ell) = 1$, within EMM)  reduces these {\it {missing moment}}, or initialization moment, variables, to two. We can represent the moment-missing moment dependence by the relations

\begin{equation}
u_\sigma(\rho) = \sum_{\ell = 0}^{m_s} M_{E,\sigma}(\rho,\ell) u_\sigma(\ell),
\end{equation}
where $M_{E,\sigma}(\rho,\ell)$ are known energy dependent polynomials (or more generally, rational fraction polynomials in $E$), satisfying the corresponding moment equation subject to the initialization conditions, $M_{E,\sigma}(\ell_1,\ell_2) = \delta_{\ell_1,\ell_2}$.

\subsection{  Orthogonal Polynomial Projection Quantization}
We review the Orthogonal Polynomial Projection Quantization method in general, and then apply it to QES systems.

Suppose ${\cal A}(x) > 0$ is a positive, bounded, configuration admitting an infinite set of orthonormal polynomials (our bra-ket notation will omit explicit reference to the underlying weight, for simplicity)

\begin{eqnarray}
\langle {\cal P}^{(j_1)}| {\cal P}^{(j_2}\rangle \equiv \int dx {\cal P}^{(j_1)}(x) {\cal P}^{(j_2)}(x) {\cal A}(x) =  \delta_{j_1,j_2}, \nonumber \\
{\cal P}^{(j)}(x) = \sum_{i = 0}^j \Xi_i^{(j)} x^i,  \ where \ \Xi_j^{(j)} \neq 0.
\end{eqnarray}
Assume that the quantum system under consideration admits a moment equation, represented as

\begin{equation}
\mu(p) = \sum_{\ell = 0}^{m_s} M_E(p,\ell) \ \mu(\ell), p \geq 0.
\end{equation}

Consider expanding the desired discrete state in terms of the orthonormal polynomial basis:

\begin{eqnarray}
\Psi(x) = \sum_{j=0}^\infty \Omega_j {\cal P}^{(j)}(x) \ {\cal A}(x).
\end{eqnarray}
One can then project out the expansion coefficients exactly:

\begin{eqnarray}
\Omega_j & = & \int dx \ {\cal P}^{(j)}(x) \Psi(x) , \\
\Omega_j  & = & \sum_{i=0}^j \Xi^{(j)}_i \mu(i) , \\
\Omega_j & = & \sum_{\ell = 0}^{m_s} \Big( \sum_{i = 0}^j\Xi^{(j)}_i M_E(i,\ell)\Big ) \mu(\ell).
\end{eqnarray}

Now consider the positive (and assumed finite)  integral expression $\int dx {{\Psi^2(x)}\over {{\cal A}(x)}} < \infty$.
We obtain
\begin{equation}
\int dx {{\Psi^2(x)}\over {{\cal A}(x)}}= \sum_{j = 0}^\infty \Omega_j^2 < \infty,
\end{equation}
resulting in
\begin{equation}
\lim_{j \rightarrow \infty}\Omega_j = 0.
\end{equation}
The integral condition in Eq.(13) can be satisfied if ${\cal A}$ decays slower than $\Psi^2(x)$ allowing the ratio to be integrable. A rougher statement suggests that if ${\cal A}$ decreases as, or  slower, than $\Psi$ then the above is satisfied. Note that we do not want ${\cal A}$ decreasing so fast that even the unphysical solutions have finite integrals. In this case, the OPPQ quantization conditions will not work. These considerations are also essential within the EMM analysis. 

The asymptotic behavior of Eq.(14) suggests that we impose these conditions, at finite order, on appropriate, successive, projection expressions as represented in Eq.(12).
In particular:

\begin{eqnarray}
 \sum_{\ell = 0}^{m_s} \Big( \sum_{i = 0}^j\Xi^{(j)}_i M_E(i,\ell)\Big ) \mu(\ell) = 0,
\end{eqnarray}
for $j = N,N+1,N+2,\ldots, N+m_s$, defining an $(m_s+1)\times (m_s+1)$ determinantal condition:

\begin{eqnarray}
D_N(E) = Det( {\cal M}_{\eta,\ell}^{(N)}(E) ) = 0,
\end{eqnarray}
where ${\cal M}_{\eta,\ell}^{(N)}(E) = \sum_{i = 0}^{N+\eta}\Xi^{(N+\eta)}_i M_E(i,\ell)$. We note that the degree of $D_N(E)$, generally a rational polynomial of the energy , grows as $N \rightarrow \infty$, allowing for the generation of converging approximants to all the discrete states.

The energy roots to Eq.(16) generally converge, exponentially fast to the physical energies. The closer ${\cal A}$ describes the asymptotic behavior of the desired physical state, the faster the convergence.  We emphasize that Eq.(16) is valid only if there is no symmetry related condition requiring the any of the missing moments  be zero. For parity invariant systems, the above formalism should be implemented within the moment representation for that symmetry class (i.e. Eq.(5)).

Several important features distinguish OPPQ with respect to other methods. First of all, with respect to determining the energies, one does not need the explicit form for ${\cal A}$. All that is required is that one be able to generate the orthogonal polynomials accurately. Furthermore, this asymptotic factor does not have to be differentiable.
Indeed, for the quartic potential, $V(x) = x^4$, using ${\cal A}(x) = exp(-{|x|^3}\over 3)$ gives better results than using the gaussian.

\subsection{ Generating the Orthonormal Polynomials for ${\cal A}$}

This subsection is included for completeness. The orthonormal polynomials of ${\cal A}(x)$ can be determined through the three term recursion relation for their monic form. Let us denote the monic polynomials by ${\tilde{\cal P}}^{(j)}(x) = {1\over {\Xi_j^{(j)}}} {\cal P}^{(j)}(x)$.  For simplicity, we shall refer to the leading orthonormal coefficient as $n_j \equiv \Xi_j^{(j)}$.

Let ${\tilde{\cal P}}^{(j)}(x) = {1\over {n_j}} {\cal P}^{(j)}(x) = x^j+b_j x^{j-1} + \ldots$, represent the monic form of the orthonormal polynomial, ${\cal P}^{(j)}(x)$. We then have  $\langle \tilde{\cal P}^{(j)}| \tilde{\cal P}^{(j)}\rangle = \langle x^j|\tilde{\cal P}^{(j)}\rangle$, and $\langle x \tilde{\cal P}^{(j)}| \tilde{\cal P}^{(j)}\rangle = \langle x^{j+1}| \tilde{\cal P}^{(j)}\rangle + b_j  \langle x^j|\tilde{\cal P}^{(j)}\rangle$. 
The monic orthogonal polynomials  satisfy the well known three term recurrence relation, 
\begin{eqnarray}
{\tilde{\cal P}}^{(j+1)}(x) = (x-{\tilde \alpha}_{j+1}) {\tilde{\cal P}}^{(j)}(x) - {\tilde\gamma}_j {\tilde{\cal P}}^{(j-1)}(x)
\end{eqnarray}
for$ j \geq 0 $, where ${\tilde{\cal P}}^{(-1)}(x) \equiv 0$,${\tilde{\cal P}}^{(0)}(x) \equiv 1$, and 
${\tilde\alpha}_{j+1} = {\langle x{\tilde{\cal P}}^{(j)}  |   {{\tilde{\cal P}}^{(j)}\rangle}\over {\langle x^j|{\tilde{\cal P}}^{(j)}\rangle}}$ and ${\tilde\gamma}_j = {\langle x^j|{\tilde{\cal P}}^{(j)}\rangle}\over {\langle x^{j-1}|{\tilde{\cal P}}^{{(j-1)}}\rangle}$. 

All these expressions depend on the  power moments of the weight $m(p) = \int dx x^p {\cal A}(x)$, which are  assumed  known. Specifically $\langle x {\tilde{\cal P}}^{(j)}|{\tilde{\cal P}}^{(j)}\rangle = \sum_{i_1 = 0}^j\sum_{i_2 = 0}^j  \Xi^{(j)}_{i_1} \Xi_{i_2}^{(j)} m(i_1+i_2+1)$, 
$\langle x^j|{\tilde{\cal P}}^{(j)}\rangle = \sum_{i = 0}^j   \Xi_{i}^{(j)} m(i+j)$, and $\langle x^{j-1}|{\tilde{\cal P}}^{(j-1)}\rangle = \sum_{i = 0}^{j-1}   \Xi_{i}^{(j-1)} m(i+j-1)$.
 Given the monic orthogonal polynomials, $\{{\tilde{\cal P}}^{(j)}| j \leq J \}$, the moments $\{m(2j+1),m(2j),m(2j-1),\ldots,m(0)| j \leq J\}$ are required for generating the ${\tilde \alpha}_{J+1}$ and ${\tilde \gamma}_J$ coefficients for generating the next monic orthogonal polynomial, ${\tilde P}^{(J+1)}(x)$.  The three term recursion relation is usually  the preferred procedure for generating the monic orthogonal polynomials. The coefficient $n_j$ is then obtained from $n_j^2 \langle {\tilde{\cal P}}_j|{\tilde{\cal P}}_j \rangle = 1$.  That is ${\tilde{\gamma}}_j = ({{n_{j-1}}\over{n_{j}}})^2$, involving the ratio of the norms.

We can transform the monic three term recursion relation into the counterpart for the orthonormal polynomials:

\begin{eqnarray}
{{\cal P}}^{(j+1)}(x) = (x-{\tilde \alpha}_{j+1}) \rho_j{{\cal P}}^{(j)}(x) - {\tilde\gamma}_j \rho_{j} \rho_{j-1}{{\cal P}}^{(j-1)}(x),
\end{eqnarray}
for$ j \geq 0 $, where ${\tilde{\cal P}}^{(-1)}(x) \equiv 0$, and 
$\rho_j = {{n_{j+1}}\over {n_j}}$.

An alternative representation for the orthogonal polynomials comes from Pade analysis [20] which yields

\begin{eqnarray} 
{\tilde{\cal P}}^{(j)}(x)  = {1\over {{\Delta}_{0,j-1}(m)}} Det \pmatrix {& m(0) & m(1)& \ldots & m(j) \cr & m(1) & m(2) & \ldots & m(j+1) \cr & \ldots  & \ldots & \ldots & \ldots  \cr &  m(j-1) &  m(j) & \ldots  &m(2j-1) \cr & 1  & x & \ldots & x^j \cr}\ \nonumber \\
\nonumber \\
{\Delta}_{i,j-1}(m) = Det \pmatrix {& m(i) & m(i+1)& \ldots & m(i+j-1) \cr & m(i+1) & m(i+2) & \ldots & m(i+j) \cr & \ldots  & \ldots & \ldots & \ldots  \cr &  m(i+j-1) &  m(i+j) & \ldots  &m(i+2j-2)} \nonumber > 0, \ for \ i = 0,1.\\
\end{eqnarray}
The latter correspond to the Hankel-Hadamard determinants, which must be positive for a (non-atomic) nonnegative weight (although OPPQ requires ${\cal A}$ to be positive). Note then that $\langle {\tilde{\cal P}}^{(j)}|{\tilde{\cal P}}^{(j)} \rangle = \langle x^j|{\tilde{\cal P}}^{(j)}\rangle = {{\Delta_{0,j}(m)}\over {\Delta_{0,j-1}(m)}} = n_j^{-2}$.

We can  project out, exactly, the $\Omega$ coefficients through
\begin{eqnarray}
\Omega_j = & \int_{-\infty}^{+\infty} dx \ {\cal P}^{(j)}(x) \Psi(x) ,\nonumber \\
= &{1\over \sqrt{\Delta_{0,j-1}(m) \Delta_{0,j}(m)}} Det \pmatrix {& m(0) & m(1)& \ldots & m(j) \cr & m(1) & m(2) & \ldots & m(j+1) \cr & \ldots  & \ldots & \ldots & \ldots  \cr &  m(j-1) &  m(j) & \ldots  &m(2j-1) \cr & \mu(0)  & \mu(1) & \ldots & \mu(j) \cr}. \nonumber \\
\end{eqnarray}
\vfil\break
\section{OPPQ and Quasi-Exactly Solvable Quantum Systems}

This work solely focuses on QES systems; however, for completeness, we contrast their structure with systems referred to as exactly solvable (ES), for which all states are determinable in closed form. In one space dimension, in some suitable coordinate transformed space if necessary, $s=s(x)$, the wavefunction for an ES system will assume the form $ \Psi(s) = s^\gamma{\cal P}^{(n)}(s) {\cal A}(s)$,
 where the positive asymptotic form is known in closed form, ${\cal A}(s) > 0$, and ${\cal P}^{(n)}(s)$  is the orthogonal polynomial relative to some positive weight ${\cal W}(s) > 0$. As before, $\gamma$ denotes any required indicial exponent.  The application of OPPQ to ES systems will be discussed in a subsequent work.

Quasi-exactly solvable (QES) systems are those admitting  wavefunctions of the form 
\begin{equation}
\Psi(x) = x^\gamma P_d(x) {\cal A}(x),
\end{equation}
 (assuming $s(x) = x$), where $P_d(x)$ is a polynomial of  degree ``$d$", to be determined, and not necessarily the orthogonal polynomial of any weight.  Within  OPPQ, such states will have exactly solvable energies and wavefunctions (i.e. the roots of closed form algebraic functions of the energy, etc.). This statement implicitly assumes the existence of a moment equation. For the remainder of the subsequent presentation (i.e. the sextic anharmonic oscillator), we will take $\gamma = 0$.

From the discussion and definitions in the previous sections, since $P_d(x) = \sum_{j=0}^d c_j {\cal P}^{(j)}(x) $, we know that for the QES states:
\begin{eqnarray}
\int dx {\cal P}^{(j)}(x)\Psi(x)  = 0,  \ j \geq d+1.
\end{eqnarray}
That is, the OPPQ quantization condition in Eq.(14) is exactly satisfied at all orders above a certain threshhold ($j \geq d+1$).
Assuming that the corresponding missing moments are not identically zero (for that particular state), then Eq.(15) is satisfied for all $N \geq d+1$. Since the $M_{E}(i,\ell)$ expressions are known in closed form (usually producing an algebraic function of $E$ for the determinant in Eq.(16)) it means that the discrete state energy would be determined in closed form, as the {\it constant} roots of Eq.(16) for all orders $N \geq d+1$. That is, for QES systems, the determinant quantization expression in Eq.(16) will admit two types of roots, for $N \geq d+1$. There will be the varying roots that  converge (exponentially fast) to the true, non-QES states, of the system. The other roots, for arbitrary $N \geq d+1$ will not vary and correspond to the exact energies. 

Upon determining the QES energies, the corresponding missing moment values are determined,  thereby yielding the OPPQ projection coefficients (i.e. $\Omega_j$'s), thereby generating the wavefunction.

\subsection{Additional Moment Identities for  QES Solutions}

Although OPPQ is dependent on the existence of  a moment equation, there is another moment relation inherent to  QES solutions  that is independent of the existence of such moment equations, but strongly suggest that these  systems must admit some form of moment equation.

 If we take ${\cal P}^{(j)}(x) = \sum_{i = 0}^j \Xi_{i}^{(j)} x^{i}$, where $\Xi_j^{(j)} \neq 0$, and insert in Eq.(22), or $\langle {\cal P}^{(j)}|\Psi\rangle = 0$, for $j \geq d+1$, we obtain:

\begin{eqnarray}
\mu(j) = -{1\over {\Xi_j^{(j)}}}\sum_{i = 0}^{j-1}\Xi_i^{(j)} \mu(i) , j \geq d+1.
\end{eqnarray}
In particular, starting at $j = d+1$, this linear, recursive, relation connects all the moments $\{\mu(j)| j\geq d+1\}$ to the lower order moments $\{\mu(j)| j \leq  d\}$. These relations are  not valid for the non-QES states, since $\langle {\cal P}^{(j)}|\Psi_{non-QES}\rangle \neq 0$. If the system in question has a moment equation, represented as $\mu(p) = \sum_{\ell = 0}^{m_s} M_E(p,\ell) \mu(\ell)$, $p \geq 0$, then for $p \geq d+1$, the moment equation and Eq.(23) must yield the same results once the QES energy and corresponding missing moments have been determined.

If the system is parity invariant, the orthonormal polynomials will involve polynomials of alternating even degrees and odd degrees. Therefore, for the even states, if $d = 2n$, then Eq.(23) holds for $j = d+2,d+4,\ldots$. If $d = 2n+1$, then $j =d+2,d+4,\ldots$. 

\subsection{QES-OPPQ Analysis of Sextic Anharmonic Oscillator}

Consider the sextic anharmonic oscillator problem in Eq.(1). The leading asymptotic form for the physical bound states corresponds to 
\begin{equation}
{\cal A}(x) = exp\Big(  -{{\sqrt{g}}\over 4}\big(x^4 +{b\over g} x^2\big)\Big).
\end{equation}
We will illustrate the  consistency of the  OPPQ analysis applied within the $\Psi$- representations (i.e. OPPQ-$\Psi$), which corresponds, in this case, to an $m_s=2$ moment equation representation. However, the ideal representation that recovers the Bender-Dunne energy polynomials is that defined by $\Phi(x) = {\cal A}(x) \Psi(x)$, an $m_s = 0$ problem, as discussed in the next section, and referred to as the OPPQ-$\Phi$ analysis.  

As will be seen in the next section, within the $\Phi$ representation, the particular form of the corresponding moment equation will readily reveal the existence of QES states. Within the $\Psi$ representation, this becomes more difficult, unless one specifically implements a Hill representation analysis and confirms the truncation of the $A(x)$ power series factor. However, a systematic examination of the JWKB form for the wavefunction can suggest the possible existence of QES states.

A simple, first order, JWKB approximation for the sextic anharmonic problem suggests that there are QES discrete state wavefunctions  of the form $\Psi_\sigma(x) = P_d(x) {\cal A}(x)$, where  $d = 2n_*+\sigma_*$, and $\sigma_* = 0 \ or \ 1$, for the even or odd states, respectively.
More specifically, the first order JWKB asymptotic form of the discrete state wavefunction gives $\Psi(x) \sim {1\over {({\partial_xS(x))^{1\over 2}}}} exp(-S(x))$, where 
\begin{equation}
S(x) = {\sqrt{g}\over 4}(x^4 +{b\over g} x^2) + {1\over {\sqrt{g}}} ({m\over 2} -{{b^2}\over{8g}})Ln(x).
\end{equation}
 The asymptotic estimate becomes  $\Psi_{\sigma}(x) \sim  x^{d} {\cal A}(x)$, where 
$d  = {1\over {\sqrt{g}}} ( {{b^2}\over{8g}}-{m\over 2})-{3\over 2}$. Since there can only be even or odd solutions, the potential function parameters leading to an integer form for $d = 2n_*+\sigma_*$ correspond to the QES potential function constraints in Eq.(2). We can test the validity of this by checking that the OPPQ analysis yields constant QES energy values within the OPPQ framework. As previously noted, the constraint in Eq.(2) can only be explicitly confirmed either within a Hill representation (truncation) analysis, or the $\nu$-moment analysis in the next section.

We will work within each parity symmetry class associated with the QES state. The QES form for the wavefunction will be $\Psi_{\sigma_*}(x) = P_d(x) {\cal A}(x)$ , where $P_d(x) \rightarrow  x^{\sigma_*} { P}_{n_*}(x^2) $,  $d = \sigma_*+2n_*$, for the even or odd  states ($\sigma_* = 0,1$): 

\begin{equation}
\Psi_{\sigma_*}(x) = x^{\sigma_*} { P}_{n_*}(x^2)  {\cal A}(x).
\end{equation}
For notational simplicity, the following discussion implicitly assumes that all references to $\sigma,n$ implicitly refer to the QES values $\sigma_*,n_*$.
 We  expand the wavefunction in terms of

\begin{equation}
\Psi_{\sigma}(x) = \sum_{j=0}^n \Omega_j x^{\sigma} {{\cal P}}^{(j)}_{\sigma}(x^2) {\cal A}(x),
\end{equation}
 where $x^{\sigma}{\cal P}^{(j)}_{\sigma}(x^2)$ are the even and odd orthonormal polynomials of ${\cal A}$, satisfying $\langle {{\cal P}}_{\sigma}^{(j_1)} | x^{2{\sigma}} {\cal A}(x) |{{\cal P}}_{\sigma}^{(j_2)} \rangle = \delta_{j_1,j_2}$. 

Quantization via OPPQ involves 

\begin{eqnarray}
\int dx\  x^\sigma{{\cal P}}^{(j)}_\sigma(x^2) \Psi_\sigma(x) = 0, \nonumber\\
\int_0^\infty d\xi  \ {{\cal P}}^{(j)}_\sigma(\xi)\  \psi_\sigma(\xi) = 0, \ for \  j \geq n_*+1,
\end{eqnarray}
 where  $\psi_\sigma(\xi) \equiv \xi^{{\sigma-1}\over 2} \Psi_\sigma(\sqrt{\xi})$,  from Eq.(4). 

Writing ${{\cal P}}^{(j)}_\sigma(\xi) = \sum_{i=0}^j \Xi_{\sigma;i}^{(j)}\  \xi^i $,  Eq.(27) transforms into 

\begin{eqnarray}
\sum_{i=0}^j \Xi_{\sigma;i}^{(j)}\  u_\sigma(i) = 0, \nonumber \\
\sum_{\ell = 0}^{m_s} \Big(  \sum_{i = 0}^j\Xi_{\sigma;i}^{(j)} M_{E,\sigma}(i,\ell)\Big ) u_\sigma(\ell) = 0, \ for \ j \geq n_*+1,
\end{eqnarray}
using the $u_\sigma$-moment equation in Eq.(6).  As suggested in Eq.(29), this relation is exactly true for QES states. It becomes the OPPQ approximation for non-QES states.

 The missing moment order is $m_s = 2$, therefore Eq.(29) must be valid for any three successive $j$ values, provided they are greater than $n_*+1$. In particular, for $ j = N,N+1, N+2$, where $N \geq n_*+1$, we obtain the determinantal relation
\flushleft
\begin{eqnarray}
D_N(E) = Det \pmatrix{ {\cal M}_{(N,0)}(E) &  {\cal M}_{(N,1)}(E)  & {\cal M}_{(N,2)} (E)  \cr {\cal M}_{(N+1,0)}(E) &  {\cal M}_{(N+1,1)}(E)  & {\cal M}_{(N+1,2)} (E)  \cr {\cal M}_{(N+2,0)}(E) &  {\cal M}_{(N+2,1)}(E)  & {\cal M}_{(N+2,2)} (E)   } = 0, \ for \ N \geq n_*+1,\nonumber \\
\end{eqnarray}
where ${\cal M}_{(N+\ell_1,\ell_2)}(E) \equiv  \sum_{i = 0}^{N+\ell_1}\Xi_{\sigma;i}^{(N+\ell_1)} M_{E,\sigma}(i,\ell_2)$, $0 \leq \ell_{1,2} \leq 2$.

As stated before, the degree of the $D_N(E)$ polynomial increases with $N$.  Eq.(30) will be satisfied by all QES states for fixed index $n_*$. They will be the exact  roots of Eq.(30) for all $N \geq n_*+1$. The other roots generated from Eq.(30) will approximate, and  converge (exponentially fast) to, the non-QES energies.

Once the QES energies are determined, the corresponding missing moments  are also  determined $\{u_\sigma(0), u_\sigma(1), u_\sigma(2)\}$, subject to a convenient normalization (i.e. $u_\sigma(0) = 1$).  The OPPQ expansion coefficients in Eq.(27) are then obtained through 

\begin{eqnarray}
\Omega_j = & \int dx \ x^\sigma {{\cal P}}_\sigma^{(j)}(x^2) \Psi_\sigma(x) \nonumber \\
\Omega_j = & \sum_{i = 0}^j \Xi_{\sigma;i}^{(j)} u_\sigma(i), \nonumber \\
\Omega_j = & \sum_{i = 0}^j \Xi_{\sigma;i}^{(j)} \big( \sum_{\ell = 0}^2 M_{E,\sigma}(i,\ell) u_\sigma(\ell) \big),  \ for \ j \leq n_*,
\end{eqnarray}
generating the closed form expression for the wavefunction, as given in Eq.(26).

The final component is generating the orthonormal polynomials of ${\cal A}$. Since $\langle {{\cal P}}_\sigma^{(j_1)} | \xi^{\sigma}{{ {\cal A}(\xi)}\over{\sqrt{\xi}}} |{{\cal P}}_\sigma^{(j_2)} \rangle = \int_0^\infty d\xi \ {\cal P}_\sigma^{(j_1)}(\xi){\cal P}_\sigma^{(j_2)}(\xi) \xi^{\sigma-{1\over 2}} {\cal A}(\xi) = \delta_{j_1,j_2}$, the respective orthonormal polynomials are generated by different weights in the $\xi$-coordinate. We need the power moments of these different weights, $m_\sigma(\rho) = m(\rho+\sigma) = \int_0^\infty d\xi \ \xi^{\rho+\sigma}  {{ {\cal A}(\xi)}\over{\sqrt{\xi}}}$.

Anticipating the needs of the OPPQ-$\Phi$ representation, we define ${\cal A}(s;x) =   exp\big(  -{{\sqrt{g}}\over s}\big(x^4 +{b\over g} x^2\big)\Big)$, where  $s = 4$ for OPPQ-$\Psi$ (i.e. the  OPPQ-$\Phi$ works with ${\cal A}^2(x)$, thus requiring $s = 2$).  Since $\partial_x{\cal A}(s;x) = -{ {\sqrt{g}}\over s}\big(4 x^3 + {{2b}\over {{g}}} x\big) {\cal A}(s;x)$, this generates the moment equation (upon multiplying both sides by $x^{2\rho+1}$ and integrating by parts):
\begin{eqnarray}
m(\rho+2) = {s{(2\rho+1)}\over {4\sqrt{g}}} m(\rho) - {{b}\over {2{g}}} m(\rho+1), \rho \geq 0,
\end{eqnarray}
where $m(\rho) = \int_{-\infty}^{+\infty} dx   x^{2\rho} {\cal A}(s;x)$. We can use Mathematica to determine the $m(0)$ and $m(1)$ moments in terms of 
the modified Bessel function of the second kind $\int dx {\cal A}(x) = ({e\over 2})^{1\over 4} K_{1\over 4}({1\over 4})$, and the Bessel function of the first kind, 
$\int dx x^2{\cal A}(x) = -{\pi\over 2}({e\over 2})^{1\over 4} \Big ( I_{-{1\over 4}}({1\over 4}) -3  I_{{1\over 4}}({1\over 4}) + I_{{3\over 4}}({1\over 4}) -  I_{{5\over 4}}({1\over 4})\Big)$.

Consider the potential function parameters $g = 1$, $b^2 = 8$, then the potential function parameter becomes $m_{pot} = -(4n_*+2\sigma_*+1)$.    Tables  1 and 2  give the OPPQ analysis for the corresponding QES and non-QES states for $n_* = 3$. It will be noted that as soon as $N \geq n_*+1$, the QES states are exactly determined and remain the same constant roots for the corresponding $D_N(E)$ function. The other OPPQ energy roots for $D_N(E)$ converge to the non-QES states. We emphasize that the numbers given for the QES states represent the first six-seven decimal places of the exact energies with no rounding off. We also give the OPPQ estimate for the non-QES states, derived from a higher order OPPQ analysis using orthonormal polynomials of $exp(-{{x^4}/4})$ as developed in Ref. [11].

 For completenss, Tables 3 and 4 give both QES and non-QES states derived without working in the explicit parity subspaces. That is, we work with the $\mu$ moments directly ($m_s = 5$), generating the corresponding ($6 \times 6$) OPPQ determinantal equation. The $N$ paramter quoted is different from that in Talbes 1 and 2. For Tables 3 and 4,  for the $n_* = 3$ case, the QES states have $P_d(x)$ with $d = 2n_*+\sigma_*$, hence the exact QES energies become manifest for $N \geq 7 \ or \ 8$, depending on the even or odd states, respectively.

\begin{table}
\caption{\label{table2}
Convergence of OPPQ-$\Psi$  (QES* and non-QES) for the first six (even) energy levels of Eq.(1), $g=1$,$b=\sqrt{8}$, $m = -(4n_*+2\sigma_*+1)$,$n_* = 3$, $\sigma_* = 0$ }
\begin{tabular}{ccccccc}
$N$ & $E_0^*$ & $E_2^*$ & $E_4^*$ & $E_6^*$ & $E_8$ & $E_{10}$\\\hline
\multicolumn{7}{c}{\vrule height14pt depth5pt width0pt  $V(x) = gx^6+bx^4+mx^2$, ${\cal A}(x) =   exp\big(  -{{\sqrt{g}}\over 4}\big(x^4 +{b\over g} x^2\big)\Big)$}\\
\hline
1 & -3.500501 &   &  &  & &  \\  
2 & -6.604075 & 0.507807   &  &  &  &   \\ 
3 &  -4.538891& 2.361563   &  8.006481  &  &  &    \\ 
4 & -4.701631 & 2.289850  & 13.186912  &28.822848 & &  \\ 
5 & -4.701631 & 2.289850  & 13.186912  & 28.822848 & 61.179448 &   \\  
6 & -4.701631 & 2.289850  &13.186912 & 28.822848 & 51.599563 &  102.816240  \\  
7 &-4.701631& 2.289850  & 13.186912  & 28.822848 & 48.712815 &  82.421165   \\ 
8 & -4.701631 & 2.289850   & 13.186912  & 28.822848 & 47.857837& 74.249292  \\ 
9 & -4.701631 & 2.289850  & 13.186912  &28.822848 & 47.652156& 70.817047  \\ 
10 & -4.701631 & 2.289850  & 13.186912 & 28.822848 & 47.616909 & 69.545484   \\  
11& -4.701631 & 2.289850  & 13.186912  &28.822848 &47.614022 & 69.227821 \\ 
12 & -4.701631 & 2.289850& 13.186912 & 28.822848 & 47.613850 & 69.232777   \\
$\infty$
  &            &        &            &           & 47.613209 & 69.043247\\ 
\end{tabular}
\end{table}

\begin{table}
\caption{\label{table2}
Convergence of OPPQ-$\Psi$  (QES* and non-QES) for the first six (odd) energy levels of Eq.(1), $g=1$,$b=\sqrt{8}$, $m = -(4n_*+2\sigma_*+1)$,$n_* = 3$, $\sigma_* = 1$ }
\begin{tabular}{ccccccc}
$N$ & $E_1^*$ & $E_3^*$ & $E_5^*$ & $E_7^*$ & $E_9$ & $E_{11}$\\\hline
\multicolumn{7}{c}{\vrule height14pt depth5pt width0pt  $V(x) = gx^6+bx^4+mx^2$, ${\cal A}(x) =   exp\big(  -{{\sqrt{g}}\over 4}\big(x^4 +{b\over g} x^2\big)\Big)$}\\
\hline
1 & -8.086559 &   &  &  & &  \\  
2 & -7.931590 &-0.067843   &  &  &  &   \\ 
3 &  -6.466044& 5.685330   &  10.297850 &  &  &    \\ 
4 & --6.629227 & 4.618850  & 18.024593 &34.897472 & &  \\ 
5 & -6.629227  & 4.618850  & 18.024593 & 34.897472& 70.431224 &   \\  
6 & -6.629227 & 4.618850  &18.024593 & 34.897472 & 59.527051 &  114.774703 \\  
7 &-6.629227& 4.618850  & 18.024593  & 34.897472 & 56.100032 & 92.408733   \\ 
8 & -6.629227 & 4.618850 & 18.024593 & 34.897472 & 55.025448& 83.209865  \\ 
9 & -6.629227& 4.618850 & 18.024593  &34.897472& 54.745576& 79.194992  \\ 
10 &  -6.629227 & 4.618850  & 18.024593 & 34.897472 & 54.689889 & 77.548343   \\  
$\infty$ &                 &                 &                      &                  & 54.686459 & 76.977398 \\
\end{tabular}
\end{table}
\vfill\break
\newpage
\begin{table}
\centering
\caption{\label{table1}
Unified OPPQ-$\Psi$ analysis within $\mu$ representation for $\sigma_*= 0, n_*= 3$,  $b=\sqrt{8}$ and $m=-13$.
QES states $E_0$, $E_2$, $E_4$ and $E_6$ are ``exact", while the other states converge fast.}
\vskip10pt\centering
\rotatebox{90}{
\small
\begin{tabular}{ccccccccc}
N & $E_0^*$ & $E_1$ & $E_2^*$ & $E_3$ & $E_4^*$ & $E_5$ & $E_6^*$ & $E_7$ \\
\hline
 1 & -3.50050190677 &   &   &   &   &   &   &   \\
 2 & -6.03169311724 & -3.50050190677 &   &   &   &   &   &   \\
 3 & -6.60407548295 & -6.03169311724 & 0.507807963155 &   &   &   &   &   \\
4 & -6.60407548295 & -4.72113209085 & 0.507807963155 & 2.25709085340 &   &   &   &   \\
 5 & -4.72113209085 & -4.53889182150 & 2.25709085340 & 2.36156311823 & 8.00648129616 &   &   &   \\
 6 & -4.53889182150 & -4.21887392229 & 2.36156311823 & 7.10373221895 & 8.00648129616 & 16.7229518999 &   &   \\
 7 & -4.70163122288 & -4.21887392229 & 2.28985002468 & 7.10373221895 & 13.1869125971 & 16.7229518999 & 28.8228483475 &   \\
 8 & -4.70163122288 & -4.25720912289 & 2.28985002468 & 6.71439942230 & 13.1869125971 & 20.2497283402 & 28.8228483475 & 43.7475563324 \\
 9 & -4.70163122288 & -4.25720912289 & 2.28985002468 & 6.71439942230 & 13.1869125971 & 20.2497283402 & 28.8228483475 & 43.7475563324 \\
 10 & -4.70163122288 & -4.25800570649 & 2.28985002468 & 6.71431004595 & 13.1869125971 & 20.5352883528 & 28.8228483475 & 39.2137668519 \\
 11 & -4.70163122288 & -4.25800570649 & 2.28985002468 & 6.71431004595 & 13.1869125971 & 20.5352883528 & 28.8228483475 & 39.2137668519 \\
 12 & -4.70163122288 & -4.25801075980 & 2.28985002468 & 6.71503370668 & 13.1869125971 & 20.5608997790 & 28.8228483475 & 38.1497221283 \\
 13 & -4.70163122288 & -4.25801075980 & 2.28985002468 & 6.71503370668 & 13.1869125971 & 20.5608997790 & 28.8228483475 & 38.1497221283 \\
 14 & -4.70163122288 & -4.25800781019 & 2.28985002468 & 6.71512397419 & 13.1869125971 & 20.5622044115 & 28.8228483475 & 37.9091389572 \\
 15 & -4.70163122288 & -4.25800781019 & 2.28985002468 & 6.71512397419 & 13.1869125971 & 20.5622044115 & 28.8228483475 & 37.9091389572 \\
 16 & -4.70163122288 & -4.25800743707 & 2.28985002468 & 6.71512989894 & 13.1869125971 & 20.5620795974 & 28.8228483475 & 37.8672556901 \\
 17 & -4.70163122288 & -4.25800743707 & 2.28985002468 & 6.71512989894 & 13.1869125971 & 20.5620795974 & 28.8228483475 & 37.8672556901 \\
 18 & -4.70163122288 & -4.25800741621 & 2.28985002468 & 6.71512975900 & 13.1869125971 & 20.5620354769 & 28.8228483475 & 37.8626239770 \\
 19 & -4.70163122288 & -4.25800741621 & 2.28985002468 & 6.71512975900 & 13.1869125971 & 20.5620354769 & 28.8228483475 & 37.8626239770 
\end{tabular}
}
\end{table}

\newpage
\vfil\break

\begin{table}
\centering
\caption{\label{table1}
 Unified OPPQ-$\Psi$ analysis within $\mu$ representation for $\sigma_*= 1, n_*=3 $,$b=\sqrt{8}$ and $m=-15$.
QES states $E_1$, $E_3$, $E_5$ and $E_7$ are ``exact", while the other states converge fast.}
\vskip10pt\centering
\rotatebox{90}{
\scriptsize
\begin{tabular}{cccccccccc}
N & $E_0$ & $E_1^*$ & $E_2$ & $E_3^*$ & $E_4$ & $E_5^*$ & $E_6$ & $E_7^*$ & $E_8$ \\
\hline
 1 & -4.31962115162 &   &   &   &   &   &   &   &   \\
 2 & -8.08655987811 & -4.31962115162 &   &   &   &   &   &   &   \\
 3 & -9.55087356079 & -8.08655987811 & -0.190781182520 &   &   &   &   &   &   \\
 4 & -9.55087356079 & -7.93159013829 & -0.190781182520 & -0.0678433862236 &   &   &   &   &   \\
 5 & -7.93159013829 & -6.51760411099 & -0.0678433862236 & 0.0554777594537 & 4.59925352308 &   &   &   &   \\
 6 & -6.51760411099 & -6.46604495681 & 0.0554777594537 & 4.59925352308 & 5.68533008405 & 10.2978504560 &   &   &   \\
 7 & -6.84097818474 & -6.46604495681 & 1.07749798410 & 5.68533008405 & 10.2978504560 & 11.7267374340 & 20.9221258143 &   &   \\
 8 & -6.84097818474 & -6.62922791805 & 1.07749798410 & 4.61885026929 & 11.7267374340 & 18.0245932316 & 20.9221258143 & 34.8974726626 &   \\
 9& -6.84901298988 & -6.62922791805 & 1.01868143168 & 4.61885026929 & 10.9341820352 & 18.0245932316 & 25.6172237107 & 34.8974726626 & 51.4874448871 \\
 10 & -6.84901298988 & -6.62922791805 & 1.01868143168 & 4.61885026929 & 10.9341820352 & 18.0245932316 & 25.6172237107 & 34.8974726626 & 51.4874448871 \\
 11 & -6.84939579745 & -6.62922791805 & 1.01724028394 & 4.61885026929 & 10.9283350058 & 18.0245932316 & 26.0186964925 & 34.8974726626 & 46.1617846405 \\
 12 & -6.84939579745 & -6.62922791805 & 1.01724028394 & 4.61885026929 & 10.9283350058 & 18.0245932316 & 26.0186964925 & 34.8974726626 & 46.1617846405 \\
 13 & -6.84941133696 & -6.62922791805 & 1.01721575801 & 4.61885026929 & 10.9291565335 & 18.0245932316 & 26.0597394744 & 34.8974726626 & 44.8467290999 \\
 14 & -6.84941133696 & -6.62922791805 & 1.01721575801 & 4.61885026929 & 10.9291565335 & 18.0245932316 & 26.0597394744 & 34.8974726626 & 44.8467290999 \\
 15 & -6.84941128633 & -6.62922791805 & 1.01721996310 & 4.61885026929 & 10.9292994586 & 18.0245932316 & 26.0625827763 & 34.8974726626 & 44.5272878301 \\
 16 & -6.84941128633 & -6.62922791805 & 1.01721996310 & 4.61885026929 & 10.9292994586 & 18.0245932316 & 26.0625827763 & 34.8974726626 & 44.5272878301 \\
 17 & -6.84941118433 & -6.62922791805 & 1.01722065544 & 4.61885026929 & 10.9293121992 & 18.0245932316 & 26.0625117133 & 34.8974726626 & 44.4658799515 \\
 18 & -6.84941118433 & -6.62922791805 & 1.01722065544 & 4.61885026929 & 10.9293121992 & 18.0245932316 & 26.0625117133 & 34.8974726626 & 44.4658799515 \\
 19 & -6.84941117246 & -6.62922791805 & 1.01722070755 & 4.61885026929 & 10.9293124794 & 18.0245932316 & 26.0624501552 & 34.8974726626 & 44.4579163980 \\
\end{tabular}
}
\end{table}

\vfil\break
 
\newpage
\section{The $\Phi$ Representation : An $m_s = 0$  Perspective on the Bender-Dunne Polynomials}

\subsection{ Transformation of  the sextic anharmonic oscillator to  an $m_s = 0$ moment equation representation}
In general, if ${\cal A}(x)> 0$ is the leading, positive, bounded, asymptotic form for the discrete wavefunction, and it is known in closed form, then if $\Phi(x) = {\cal A}(x) \Psi(x)$ admits a moment equation, it will have an order ($m_s)$  less than that in the $\Psi$ representation. For the sextic problem, this corresponds to

\begin{eqnarray}
\Phi(x) =  exp\Big(-{\sqrt{g}\over 4}\big( {{x^4}} +{b\over g} {{x^2}}\big) \Big) \Psi(x), \nonumber \\
\end{eqnarray}
whose differential equation becomes

\bigskip
\begin{eqnarray}
-\partial_x^2\Phi - \big({b\over {\sqrt{g}}} x + 2 \sqrt{g} x^3\big) \partial_x \Phi(x) \nonumber  \\
+\  \Big((m-3\sqrt{g} -{{b^2}\over{4g}})x^2 - (E+ {b\over{2\sqrt{g}}})\Big) \Phi(x) = 0.
\end{eqnarray}
Let us now assume that $\Phi(x)$ is the exponentially decaying configuration for a particular discrete state. Since it has to be continuously differentiable, we can multiply both sides of Eq.(34) by $x^p$and integrate by parts over the entire real axis. Define the power moments 
$\nu(p) = \int dx\ x^p \Phi(x)$, $p \geq 0$. These moments  satisfy the moment equation:

\begin{eqnarray}
\Big(m+3\sqrt{g} -{{b^2}\over{4g}} +2 \sqrt{g} p)\Big) \nu(p+2) = \nonumber \\
(E- {b\over{\sqrt{g}}}(p+{1\over 2}))\nu(p) + p(p-1)\nu(p-2), \ p \geq 0.
\end{eqnarray}
Given that the physical system admits parity invariant states, the moment equation decouples into the corresponding even and odd order power moments. 

Define  $v_\sigma(\rho) = \nu(2\rho+\sigma)$, $\sigma = 0,1$, corresponding to the even and odd states, respectively.  The corresponding moment equations becomes:

\begin{eqnarray}
\Big(m+3\sqrt{g} -{{b^2}\over{4g}} +4 \sqrt{g} \rho)\Big) v_e(\rho+1) = \nonumber \\
(E- {b\over{\sqrt{g}}}(2\rho+{1\over 2}))v_e(\rho) + 2\rho(2\rho-1) v_e(\rho-1), \ \rho \geq 0;
\end{eqnarray}
and
\begin{eqnarray}
\Big(m+3\sqrt{g} -{{b^2}\over{4g}} +2 \sqrt{g} (2\rho+1))\Big) v_o(\rho+1) = \nonumber \\
(E- {b\over{\sqrt{g}}}(2p+{3\over 2}))v_o(\rho) + 2\rho(2\rho+1) v_o(\rho-1), \ \rho \geq 0.
\end{eqnarray}
We can express the above more compactly as
\begin{eqnarray}
\Big(m -{{b^2}\over{4g}} + \sqrt{g}(4 \rho + 3 + 2\sigma)\Big) v_{\sigma}(\rho+1) = \nonumber \\
(E- {b\over{\sqrt{g}}}(2\rho+{1\over 2}+\sigma))v_{\sigma}(\rho) + 2\rho(2\rho-1+2 \sigma) v_{\sigma}(\rho-1), \nonumber\\
\end{eqnarray}
$\rho \geq 0$, and $\sigma = 0,1$.

The above moments correspond to different Stieltjes measures. Specifically, 

\begin{eqnarray}
v_{\sigma}(\rho) & =\nu(2\rho+\sigma)  = \int_{-\infty}^{+\infty} dx \ x^{2\rho+\sigma} \Phi_{\sigma}(x)  \nonumber \\
v_{\sigma}(\rho) & = \int_0^\infty d\xi \ \xi^\rho \phi_{\sigma}(\xi), 
\end{eqnarray}
where  $\ \phi_{\sigma}(\xi) \equiv  {{\Phi_{\sigma}(\xi)}\over {( {{\sqrt{\xi}}})^{1-\sigma}}}$, $ \xi \equiv x^2$, and  $\Phi_{\sigma}(x) = {\cal A}(x) \Psi_{\sigma}(x)$. 

\bigskip\noindent{\underbar{ The QES-States}}
\smallskip

The moment equations in Eq.(35-38) are implicitly only valid for the physical states. In general, except for the special QES states, they are of missing moment order $m_s = 0$ since if $v_\sigma(0) \neq 0$ (which is the case for the ground and first excited states within EMM),
all the higher order moments are generated and become polynomials in the energy:

\begin{eqnarray}
v_\sigma(\rho) = Polynomial \ of \ degree \  \rho \ in \ the\ energy, \ E ;\nonumber \\
v_\sigma(\rho) \equiv \Lambda_{\sigma}^{(\rho)}(E).
\end{eqnarray}
We see that if the coefficient of the $v_\sigma(\rho+1)$ term in Eq.(38) is never zero, for any integer $\rho$ and $\sigma$ value, then an infinite number of such polynomials are generated. 

 If this coefficient is zero for some $\rho = n_*$ and $\sigma_* = 0,1$, then the potential function parameters are constrained to
\begin{equation}
m -{{b^2}\over{4g}} + \sqrt{g}(4 n_* + 3 + 2\sigma_*) = 0,
\end{equation}
allowing only the first $n_*+1$ moments to be generated, $\{v_{\sigma_*}(\rho) | 0 \leq \rho \leq n_*\}$.  Eq.(41) is the QES parameter condition in Eq.(2). We stress that if Eq.(41) is satisfied by the potential function parameters, the states of opposite parity to the QES states, $\sigma \neq \sigma_*$, will satisfy the corresponding version of Eq.(38), and generate all the moments as polynomials in the energy $\{v_\sigma(\rho)|\rho \geq 0\}$.

Define the $n+1$ degree polynomial

\begin{eqnarray}
{ P}_{\sigma}^{(n+1)}(E) = (E- {b\over{\sqrt{g}}}(2n+{1\over 2}+\sigma))\Lambda_{\sigma}^{(n)}(E) + 2n(2n-1+2 \sigma) \Lambda_{\sigma}^{(n-1)}(E) .  \nonumber \\
\end{eqnarray}
Within the EMM framework, Handy and Bessis [14] realized that if the QES parameter conditions in Eq.(41) are satisifed, then the fact that the ground and first excited states must have nonzero, zeroth-order moments,
$v_{\sigma_*}(0) \neq 0$, makes them the roots of the respective polynomial

\begin{equation}
{ P}_{\sigma_*}^{(n_*+1)}(E) = 0.
\end{equation}
If any other excited state has its zeroth order moment also nonzero, then it too must be a root of the above polynomial. The question is, will this property also hold for all the first $n_*$ excited states? The answer is yes.
The proof, known to HB, is given below. That is, all the $n_*+1$ roots of this polynomial correspond to the QES states. 

\bigskip\noindent{\subsection{Moments' Proof that all the roots of $P_{\sigma^*}^{(n_*+1)}(E) = 0$, correspond to the QES states} }
\smallskip

We assume that the potential function parameters satisfy the constraint in Eq.(41).  Within the EMM framework,  $P_{\sigma^*}^{(n_*+1)}(E) = 0$ yields the QES energy root corresponding to the lowest energy within the $\sigma^*$ (even/odd parity) symmetry class, since the corresponding zeroth order moment is non-zero, $v_{\sigma^*}(0) \neq 0$,  due to the underlying positivity (if $\sigma_* = 0$) or nonnegativity (if $\sigma_* = 1$), for the ground or first excited state Stieltjes measure (Eq.(4)), respectively. The other higher energy states (i.e. second, third, etc.) in the $\sigma_*$- symmetry class must satisfy  Eq.(38) for the moments $\{v_{\sigma^*}(\rho)|0 \leq \rho \leq n_*\}$. Given that this is an $m_s = 0$ moment equation, there are only two possibilities for any such excited state: $v_{\sigma^*}(0) = 0$, or $v_{\sigma^*}(0) \neq 0$. If the second option holds, then that energy must be a root of the $n_*+1$ degree polynomial given in Eq.(43). Therefore, we focus on the first option, which although a real possibility, will be shown not to hold, for any of the first $n_*+1$ states (i.e. the QES states) . In fact, there are two proofs for this. We give the original (unpublished)  analysis, followed by the proof based on assuming the states have the OPPQ/QES form discussed previously: $\Psi(x) = P_d(x) {\cal A}(x)$ or $\Phi(x) = P_d(x) {\cal A}^2(x)$. 

If we assume that a particular excited state has $v_{\sigma_*}(0) = 0$, then the moment equation tells us that $v_{\sigma^*}(\rho) = 0$, for $0 \leq \rho \leq n_*$ ,although not necessarily for $v_{\sigma_*}(n_*+1)$, since it is not generated by the moment equation. 

\bigskip{\noindent{\underbar{EMM-Moment Equation Proof}}}
\smallskip

The Sturm-Liouville character of the sextic problem tells us that within the symmetry class corresponding to $\sigma^*$,  all states are uniquely characterized by the number of nodes along the positive real axis, $\xi =x^2 > 0$. The ground state has no nodes at all. The first excited state has no nodes along the positive axis (its only node is at the origin).  The next higher energy state, within the even parity or odd parity states, will have one node along the positive axis, etc. Let us denote the first  $n_*+1$ states  within the $\sigma^*$ symmetry class ( ordered in terms of energy or number of nodes on the positive real axis) by   $\phi_{\sigma^*,j}(\xi)$, $0 \leq j \leq n_*$. The lowest energy state ($j = 0$) is either the ground or first  excited state, depending on $\sigma^*$.  Let $\{ r_{\sigma^*,j;i}| 1 \leq i \leq j\}$ denote the nodes along the positive $\xi$ - real line, for the $j \geq 1$ state; therefore the configuration $\pi_{\sigma^*,j}(\xi) = \phi_{\sigma^*,j}(\xi)\Pi_{i=1}^j(\xi-r_{\sigma^*,j;i}) $ must be nonnegative (i.e. can be chosen as such). However, this means that its power moments must be positive: $\int_0^\infty d\xi \ \xi^\rho \pi_{\sigma^*,j}(\xi)  > 0$. In particular, the zeroth moment is the  linear superposition of all the first $1+j$ moments of $\phi_{\sigma^*,j}(\xi)$  (i.e. $\{v_{\sigma^*}(0),\ldots,v_{\sigma^*}(j)\}$). However, our starting assumption is that all of these are zero, provided $j \leq n_*$. This is a contradiction, so Eq.(43) is the quantization condition for all the first $n_*+1$, QES states.

\bigskip\noindent{\underbar {OPPQ-QES Proof}}
\smallskip

From OPPQ-$\Psi$ we argued that the QES states must have the form $\Psi(x) = P_d(x) {\cal A}(x)$, or $\Phi_{\sigma_*}(x) = x^{\sigma_*}P_{n_*}(x^2) {\cal A}^2(x)$. Let ${\cal O}^{(j)}_{\sigma_*}(x^2)$ denote the orthonormal polynomials of the respective even weights $x^{2\sigma_*}{\cal A}^2(x)$. We therefore have
$\int dx {\cal O}^{(n_*+q)}_{\sigma_*}(x^2) x^{\sigma_*}\Phi_{\sigma_*}(x) = 0$, for $q \geq 1$. However, these integrals correspond to a linear sum of the power moments $\{v_{\sigma_*}(0),\ldots,v_{\sigma_*}(n_*+q)\}$. If all $v_{\sigma_*}(\rho )= 0$, for $\rho \leq n_*$, then so too must $v_{\sigma_*}(n_*+1)$, and thereby all the higher order moments. This essentially would imply that $\Phi_{\sigma_*}(x) = 0$, a contradiction.

We note that both proofs rely on the existence of an $m_s = 0$ moment equation for the first $n_*+1$ moments. Neither makes use of the moment equation for the moment of order higher than $n_*$. 

\bigskip\noindent{\bf{The non-QES states of the Same Parity as the QES-State, must have $v_{\sigma_*}(\rho) = 0$, for $0 \leq \rho \leq n_*$}: $\Phi_{\sigma_*}^{(Non-QES)}(x) = \partial_x^{2n_*+2+\sigma_*}\Upsilon(x)$}

This is immediate. Since the sextic anharmonic potential is unbounded from above, there are an infinite number of bound states of either parity. If the potential function parameters satisfy Eq.(41), for some $\{\sigma_*,n_*\}$ pair, then only a finite number of the discrete states correspond to the QES states, as determined by the $n_*+1$ roots of the energy polynomial in Eq.(43). There are, therefore, an infinite number of non-QES states of the same parity as the coresponding QES states, $\sigma = \sigma_*$. These must satisfy the same moment equation as the QES states.

Only the QES states can have $v_{\sigma_*}(0) \neq 0$ because this then means that their energies are determined by Eq.(43). Therefore, the non-QES states of the same parity as the QES states must have $v_{\sigma_*}(0) = 0$, which means all the first $n_*+1$ moments are zero.  From a simple Fourier analysis, one concludes that since $\nu(p) = \int dx x^p \Phi_{\sigma_*}^{(non-QES)}(x) = 0$, for $0 \leq p \leq 2n_*+\sigma_*$ then $  \Phi_{\sigma_*}^{(non-QES)}(x)  = \partial_x^{2n_*+2+\sigma_*}\Upsilon(x)$, where $\Upsilon$ has the same parity as $\Phi_{\sigma_*}^{(non-QES)}$. This proves our claim, for the sextic anharmonic oscillator case. The same result is true for the Bender Dunne potential, with respects to the first $n_*+1$ moments being zero. However, the implications for the form of the corresponding $\Phi(x)$ configuration is complicated by the singular (indicial factor) required.

\bigskip\noindent{\underbar {Overview of the Moment Equation Structure for the QES and non-QES States}}
\smallskip

If the potential function parameters satisfy Eq.(41), we will refer to this as ``V(x) is of QES type". Unless otherwize indicated, the following discussion pertains to this case. The corresponding moment equation for the $\sigma_*$ parity states (QES or non-QES) will decouple the $v_{\sigma_*}(n_*+1)$ moment from the lower order moments. For the QES states, the first $n_*+1$ moments define an $m_s = 0$ moment equation. From Eq.(38), taking $\rho = n_*+1$, we see that the $\nu_{\sigma_*}(n_*+2)$ moment couples to $\{\nu_{\sigma_*}(n_*+1),\nu_{\sigma_*}(n_*)\}$, where the  $\nu_{\sigma_*}(n_*)$ moment, in turn,  is determined by the zeroth moment $\nu_{\sigma_*}(0)$. Thus the QES states moments' $\{\nu_{\sigma_*}(\rho)|\rho \geq n_*+2\}$ couple, linearly, to the moments  $\{\nu_{\sigma_*}(n_*+1),\nu_{\sigma_*}(0)\}$.  

In summary, the first $n_*+1$ moments, for the QES states, satisfy an $m_s = 0$ moment equation. The higher order moments will satisfy an effective $m_s =1$ moment equation.

We do not have to use the roots of  the energy polynomial in Eq.(43),  to determine the QES energies. We can apply OPPQ to the   $\{\nu_{\sigma_*}(\rho)|\rho \geq n_*+2\}$  moments ( an $m_s = 1$ moment equation). It will generate the exact QES energies for $N$ above a certain threshold. This is detailed in Sec. V.  This same OPPQ analysis will also generate  many more roots to the OPPQ determinant. These will be (converging) approximants to the non-QES energies, in the $N\rightarrow \infty$ limit.  One can verify that these OPPQ solutions correspond to solutions for which the zeroth order moment vanishes asymptotically ($\lim_{N\rightarrow \infty}v_{\sigma_*}(0) = 0$) corresponding to the non-QES states. This is also discussed in Sec. V.

Continuing with the case of ``V(x) of QES type",  the non-QES states of the same symmetry as the QES states  must satisfy the same moment equation. From Eq.(38), if $\rho = n_*+1$, then $v_{\sigma_*}(n_*+2)$ is determined by $v_{\sigma_*}(n_*+1)$ since $v_{\sigma_*}(n_*) = 0$, for these non-QES states. In other words, the non-QES states of the same symmetry as the QES states,  satisfy an $m_s = 0$ missing moment relation, with respect to the moments of order $n_*+1$ and higher. They are all linearly dependent on $v_{\sigma_*}(n_*+1)$.  Tables 5 and 6  uses OPPQ on the higher order moments to compute the non-QES states, of the same parity as the QES states, assuming $\{v_{\sigma_*}(\rho) = 0|0\leq \rho \leq n_*\}$. The details of this analysis are also  given in Sec. V. 

If V(x) is of ``QES type", then there will be non-QES states of opposite parity to the QES states. In this case, Eq.(38) is a full $m_s = 0$ moment equation, for all moments $\{v_\sigma(\rho)| \rho \geq 0\}$. 

If V(x) is not of ``QES type", then all states satisfy Eq.(38) which is, again, an $m_s = 0$, moment equation.

We summarize all the above in Table 7.

\vfil\break
\begin{table}
\caption{\label{table1}
OPPQ-$\Phi$ determination of  non-QES states (of $\sigma_*$ symmetry)  computed by taking $v_{\sigma_*}(\rho) = 0$, $0 \leq \rho \leq n_*$, and $\{v_{\sigma_*}(\rho)|\rho \geq n_*+1\}$ satisfy an effective $m_s = 0$ moment equation.
Refer to Eq.(38).}
\centerline{
\begin{tabular}{lcccc}
$N $  & $E_8$ & $E_{10}$ & $E_{12}$ & $E_{14}$      \\
\hline
\multicolumn{5}{c}{\vrule height14pt depth5pt width0pt $V(x) = x^6 + \sqrt{8}x^4-13 x^2$,  $n_*=3,\sigma_*=0$} \\
\hline
 5 &48.394656&                       &   &        \\
 6 & 47.671288&  72.503581  &                         &     \\
 7 & 47.617135&  69.537633  &  101.036761   &          \\
 8 &47.613461& 69.101670 &  94.571123&133.727067   \\
 9 & 47.613226&  69.049273  & 93.118330 &  122.972492   \\
 10 & 47.613211  & 69.044300  & 92.864698  &    119.986303 \\
11&47.613211 &69.044199  & 92.856805 & 119.850391  \\
 \hline
\end{tabular}
}
\end{table}

\begin{table}
\caption{\label{table1}
OPPQ-$\Phi$ determination of  non-QES states (of $\sigma_*$ symmetry)  computed by taking $v_{\sigma_*}(\rho) = 0$, $0 \leq \rho \leq n_*$, and $\{v_{\sigma_*}(\rho)|\rho \geq n_*+1\}$ satisfy an effective $m_s = 0$ moment equation.
Refer to Eq.(38).}
\centerline{
\begin{tabular}{lcccc}
$N $  & $E_9$ & $E_{11}$ & $E_{13}$ & $E_{15}$      \\
\hline
\multicolumn{5}{c}{\vrule height14pt depth5pt width0pt $V(x) = x^6 + \sqrt{8}x^4-15 x^2$,  $n_*=3,\sigma_*=1$} \\
\hline
 5 & 55.531291&                       &   &        \\
 6 & 54.750410&  80.668061  &                         &     \\
 7 & 54.690840&  77.512874  &  110.194217   &          \\
 8 & 54.686737&  77.041110 &  103.388049 &143.814626   \\
 9 & 54.686468&  76.982602  & 101.807743  &  132.398420    \\
 10 & 54.686447 &  76.974990  & 101.398734  &   127.471438 \\
 \hline
\end{tabular}
}
\end{table}

 \vfil\break
\begin{table}
\caption{\label{table1}
Moment Equation Structure for Sextic Anharmonic Potential for V(x) = ``QES Type''  (QES) or ``not of QES Type'' (N-QES); $\Phi_\sigma(x) = $ QES or N-QES.
 Refer to Eq.(38).}
\centerline{
\begin{tabular}{lcccccc}
$V(x) $  & $\Phi_\sigma$ & $\Phi_\sigma(x)$-type & $v_\sigma(\rho)$ & $m_s$ & $ v_\sigma(\rho)  $ & $m_s $     \\
\hline
\multicolumn{7}{c}{\vrule height14pt depth5pt width0pt } \\
N-QES  & $\sigma = 0,1$&   N-QES &      $\{v_\sigma(\rho)|\rho \geq 0\} $    &    0     &           \\
QES    &$\sigma \neq \sigma_*$&   N-QES &   $\{v_\sigma(\rho)|\rho \geq 0\} $ &  0      &   &              \\
QES    &$\sigma =  \sigma_*$&  N-QES&   $\{v_{\sigma_*}(\rho)|\rho \geq n_*+1\}^a $ &  0      & $\{v_{\sigma_*}(\rho)= 0 |\rho \leq n_* \}^a$ &          \\
QES    &$\sigma =  \sigma_*$ &  QES &   $\{v_{\sigma_*}(\rho)|\rho \geq n_*+1\}^b $ &  1      &  $\{v_{\sigma_*}(\rho) \neq 0 |\rho \leq n_* \}^b$  &    0          \\
\hline 
\end{tabular}
} $^a$Application of OPPQ recovers non-QES energies given in Tables 5 and 6. \\ $^b$ Application of OPPQ (instead of Eq.(43)) gives exact QES energies, $N \geq d+1$; and the non-QES energies in the $N\rightarrow \infty$ limit. 
\end{table}

\vfil\break

\bigskip\noindent{\subsection{Defining the 3-Term Recursive Relation for the Bender-Dunne Energy-Polynomials}}
\smallskip

For the $m_s = 0$ cases indicated in Table 7, the indicated $v_\sigma(\rho)$ moments are polynomials in the energy and  satisfy a three term recursion.  One can readily transform all of these cases  into the Bender-Dunne (monic) polynomials, although the more interesting case is for the QES states, for comparison purposes within our formulation. 
 
For future reference, we define the coefficient functions in Eq.(38):

\begin{eqnarray}
C_{\sigma;1}(m,b,g;\rho+1) & = & m - {{b^2}\over {4g}} + \sqrt{g}(4\rho+3+2\sigma) , \nonumber \\
C_{\sigma;0}(E,b,g;\rho) & = & E-{b\over{\sqrt {g}}}(2\rho+{1\over 2} + \sigma), \nonumber \\
C_{\sigma:-1}(\rho-1) & = & 2\rho(2\rho-1+2\sigma). \nonumber \\
\end{eqnarray}
No potential function parameters can lead to $C_{\sigma;1} = 0$ for two sets of $(\rho,\sigma)$ values.  To do so would require $(\rho_2-\rho_1) = -{1\over 2}(\sigma_2-\sigma_1)$, which is impossible for integer differences.

Define the (non-monic) polynomials.

\begin{eqnarray}
P_{\sigma}^{(\rho)}(E) \equiv C_{\sigma;1}(m,b,g;\rho) \ v_{\sigma}(\rho).
\end{eqnarray}
If the potential function parameters satisfy $C_{\sigma;1}(m,b,g;\rho)  \neq 0$, for all $\rho$'s and $\sigma$'s, then these will satisfy the three term recursion relation:

\begin{eqnarray}
P_{\sigma}^{(\rho+1)}(E) = & {{   (E-{b\over {\sqrt{g}}}(2\rho+{1\over 2}+\sigma))}\over{C_{\sigma;1}(m,b,g;\rho)}} P_{\sigma}^{(\rho)}(E) &\nonumber \\
&+ {{2\rho(2\rho-1+2\sigma)}\over {C_{\sigma;1}(m,b,g;\rho-1)}} P_{\sigma}^{(\rho-1)}(E) ,
\end{eqnarray}
for $ 0 \leq \rho < \infty$.  If the potential function parameters satisfy the QES conditions for a particular $(n_*,\sigma_*)$ pair, then Eq.(46) is valid only for $0 \leq \rho \leq n_*$.   Based on choosing $v_{\sigma}(0) =1$, the corresponding zeroth order polynomial becomes $P_\sigma^{(0)}(E) = C_{\sigma;1}(m,b,g;0)$.  We can always choose $v_\sigma(0)$ to give us the desired normalization for $P_\sigma^{(0)}(E) $.

The three term recursion relation in Eq.(46) does not correspond to a three term relation for monic (orthogonal ) polynomials. To do so requires the modifications discussed in the context of Eq.(18). Specifically, let ${\tilde P}_\sigma^{(\rho)} = f_\rho P_\sigma^{(\rho)}$ denote the monic form. Define 
$\beta_\rho \equiv {{f_{\rho+1}}\over f_\rho}$ and  $f_{\rho+1} = C_{\sigma;1}(m,b,g;\rho) f_\rho$. Let  ${\tilde\alpha}_{\rho+1} =  {b\over{\sqrt{g}}}(2\rho+{1\over 2} + \sigma)$, and
${\tilde\gamma}_\rho = -2\rho(2\rho-1+2\sigma) \beta_\rho$. Then the monic form of Eq.(46) becomes:

\begin{equation}
{\tilde P}^{(\rho+1)}_\sigma(E) = (E - {\tilde\alpha}_{\rho+1}){\tilde P^{(\rho)}}_\sigma(E) - {\tilde \gamma}_{\rho}{\tilde P}^{(\rho-1)}_\sigma(E).
\end{equation}
Since $f_{n_*+1} = \big(\Pi_{i=0}^{n_*}C_{\sigma_*;1}(m,b,g;i)\big) f_0$, and the QES parameter conditions correspond to $C_{\sigma_*;1}(m,b,g;n_*+1) = 0$, we see that the above monic form is valid for the QES states. Also, ${\tilde{\gamma}}_{n_*+1} = 0$ which tells us that $\langle {\tilde P}^{(n_*+1)}_{\sigma_*}|{\tilde P}^{(n_*+1)}_{\sigma_*}\rangle = 0$.
\section{ Quantization of QES and Non-QES States via OPPQ-$\Phi$}

Whereas the OPPQ-$\Psi$ formulation involved an $m_s = 2$ moment equation, its structure does not change regardless of the QES or non-QES character of the solution. That is, the same computational (numerical and algebraic) procedure generates either type of state. However, in the present  OPPQ-$\Phi$ formulation, the varying nature of the missing moment order, $m_s$, as given in Table 7, results in various OPPQ representations, as detailed below.

  Let $V_{sa}(x) = gx^6+bx^4+mx^2$ be the potential function. In the first two cases given in Table 7, ($V_{sa}$ of non-QES type, or $\sigma \neq \sigma_*$ non-QES solutions for $V_{sa}$ of QES type) the moment equation is of uniform $m_s = 0$ order for $\{v_\sigma(\rho)|\rho \geq 0\}$. The resulting OPPQ determinant is $1 \times 1$, corresponding to a pure energy polynomial whose roots generate all the discrete state energies in the $N \rightarrow \infty$ limits. 

If $V_{sa}(x)$ is of QES type, then for the QES parity class, $\sigma = \sigma_*$, both non-QES and QES states have the same moment equation. However, for the non-QES states, all moments of order no greater than $n_*$ will be zero, $v_{\sigma_*}(\rho) = 0$, if $\rho \leq n_*$. The remaining moments, 
$\{v_{\sigma_*}(\rho)|\rho\geq n_*+2\}$, linearly depend on $v_{\sigma_*}(n_*+1)$. This is also an effective $m_s = 0$ relation; and the OPPQ determinant is a $1\times 1$ energy dependent polynomial. Application of OPPQ yields the non-QES energies, as shown in Tables 5 and 6. 

For the QES states, the moments $\{v_{\sigma_*}(\rho)|\rho \geq n_*+2\}$ become linearly dependent on $\{v_{\sigma_*}(n_*+1), v_{\sigma_*}(n_*)\}$ defining an $m_s = 1$ moment equation; however, since $v_{\sigma_*}(n_*)$ is linearly dependent on $v_{\sigma_*}(0) \neq 0$, the $\{v_{\sigma_*}(\rho)|\rho \geq n_*+2\}$ become linearly dependent on $\{v_{\sigma_*}(n_*+1), v_{\sigma_*}(0)\}$. We summarize the moment equation structure in Table 8. 

 \vfil\break
\begin{table}
\caption{\label{table1}
Missing Moment Structure for Sextic Anharmonic Potential for V(x) = ``QES Type''  (QES) or ``not of QES Type'' (N-QES); $\Phi_\sigma(x) = $ QES or N-QES.
 Refer to Eq.(38).}
\centerline{
\begin{tabular}{lcccc}
$V(x)$  & $\Phi_\sigma$ & $\Phi_\sigma(x)$-type & $v_\sigma(\rho)$ & $\rho \in [a,b]$  \\
\hline
\multicolumn{5}{c}{\vrule height14pt depth5pt width0pt } \\
N-QES  &$\sigma = 0,1$      &  N-QES &   $v_\sigma(\rho) = M_{E,\sigma}(\rho,0) v_\sigma(0)$  & $[0,\infty)$\\
QES    &$\sigma\neq\sigma_*$&  N-QES &   $v_\sigma(\rho) = M_{E,\sigma}(\rho,0) v_\sigma(0)$  & $[0,\infty)$\\  \hline
QES    &$\sigma =  \sigma_*$&  N-QES &   $v_{\sigma_*}(\rho) = 0 $ &  $[0,n_*]$\\  \hline
QES    &$\sigma =  \sigma_*$&  N-QES &   $v_{\sigma_*}(\rho) = M_{E,{\sigma_*}}(\rho,n_*+1) v_{\sigma_*}(n_*+1)$ &  $[n_*+1,\infty)$\\ \hline
QES    &$\sigma =  \sigma_*$&  QES   &   $v_{\sigma_*}(\rho) = M_{E,\sigma_*}(\rho,0) v_{\sigma_*}(0)$ & $[0, n_*]$ \\          \\           
QES    &$\sigma =  \sigma_*$&  QES   &   $v_{\sigma_*}(\rho) = \pmatrix{M_{E,{\sigma_*}}(\rho,n_*+1) v_{\sigma_*}(n_*+1) \cr + M_{E,{\sigma_*}}(\rho,0) v_{\sigma_*}(0) \cr}$ & $[n_*+1,\infty)$  \\
\hline 
\end{tabular}
}  
\end{table}

Within OPPQ-$\Phi$ we must work with the orthonormal polynomials of  ${\cal W}(x) \equiv {\cal A}^2(x)$, where ${\cal A}(x)$ is defined in Eq.(33). The asymptotic exponential form of all the physical states, within the $\Phi$ representation,  is given by ${\cal W}(x)$.

For the general even and odd parity states ($\sigma = 0,1$), the OPPQ representation becomes

\begin{eqnarray}
\Phi_\sigma(x) =  \sum_{j=0}^\infty \Omega_j x^\sigma{\cal O}^{(j)}_\sigma(x^2) {\cal W}(x),
\end{eqnarray}
where $x^\sigma{\cal O}^{(j)}_\sigma(x^2) $ represent the even and odd orthonormal polynomials of the weight ${\cal W}(x)$, $\langle x^\sigma{\cal O}^{(j_1)}_\sigma|{\cal W}|x^\sigma{\cal O}^{(j_2)}_\sigma\rangle =  \delta_{j_1,j_2}$. We represent them as  ${\cal O}^{(j)}_\sigma(x^2) = \sum_{i = 0}^j \Xi_{\sigma;i}^{(j)} x^{2i}$.  The expansion coefficients are given by 

\begin{eqnarray}
\Omega_j = & \int dx\   x^\sigma{\cal O}^{(j)}_\sigma(x^2) \Phi_\sigma(x) ,\nonumber \\
 \Omega_j = & \sum_{i=0}^j \Xi_{\sigma;i}^{(j)}v_\sigma(i).
\end{eqnarray}
Depending on which case is considered, as summarized in Table 8, the $\{v_\sigma(i)\}$ moments will be linearly dependent either on $v_{\sigma}(0)$ (i.e. cases 1 and 2), $v_{\sigma}(n_*+1)$ (i.e. case 3, in which $v_{\sigma_*}(\leq n_*) = 0$), or on both (for the QES states). We can represent each of these by (using the notation in Table 7 and  8)
\begin{eqnarray}
\Omega_j = & \sum_{\ell = 0}^{m_s = 0, 1} \Big( \sum_{i=0}^j \Xi_{\sigma;i}^{(j)}  M_{E,\sigma}\big(i,\ell (n_*+1)\big)\Big) v_\sigma\big(\ell (n_*+1)\big).
\end{eqnarray}
Quantization corresponds to setting $\Omega_{N+\ell}\big(v_\sigma(0),v_\sigma(m_s (n_*+1))\big) = 0$, for $0 \leq \ell \leq m_s$, and $N \rightarrow \infty$, resulting in the determinantal condition (either $1 \times 1$ or $2 \times 2$)
\begin{eqnarray}
D_N(E) = Det \Big( {\cal M}_{\ell_1,\ell_2}(E;N) \Big ) = 0,
\end{eqnarray}
where ${\cal M}_{\ell_1,\ell_2}(E,N) =  \sum_{i=0}^{N+\ell_1} \Xi_{\sigma;i}^{(N+\ell_1)} M_{E,\sigma}(i,\ell_2 (n_*+1))$, where $0 \leq \ell_{1,2} \leq m_s \ (0 \ or\ 1)$.

\bigskip\noindent\underbar{ Non-QES-Potentials}
\smallskip

For case 1 in Table 8, $m_s = 0$ and  $D_N(E)$ corresponds to the determinant of a $1 \times 1$ matrix. The OPPQ-$\Phi$ analysis will generate rapidly converging approximants to the true physical values. The results of this are not given here, but are in keeping with Tables 3 and 4.

\bigskip\noindent\underbar{Non-QES states for QES-Potentials}
\smallskip

If  the  potential function parameters satisfy the QES conditions, let  the non-QES states be represented as  $\Phi_\sigma(x)$. If $\sigma \neq \sigma_*$, then the previous case applies and the corresponding moment equation is of $m_s = 0$ form. If $\sigma = \sigma_*$, then the non-QES states must have
$v_{\sigma_*}(\rho) = 0$ for $0 \leq \rho \leq n_*$. Only the moments $\{v_{\sigma_*}(\rho)|\rho \geq n_*+1\}$ are nonzero, and satisfy an effective $m_s = 0$ relation.  One can apply OPPQ on the nonzero moments:

\begin{eqnarray}
v_{\sigma_*}(\rho) = M_{E,\sigma_*}(\rho,n_*+1) \ v_{\sigma_*}(n_*+1), \ \rho \geq n_*+1; \nonumber \\
\sum_{i=n_*+1}^{N} \Xi_{N;i}^{(N)}  M_{E,\sigma}(i,n_*+1) = 0, N \geq n_*+2,\ the \ OPPQ \ condition. \nonumber \\
\end{eqnarray}
The results are given in Tables 5 and 6.

\bigskip\noindent\underbar{QES states: Approach- I}
\smallskip

When the potential function parameters do satisfy the QES constraints, then the  QES state energies are determined from Eq.(43).   The $\{v_{\sigma_*}(\rho)| 0 \leq \rho \leq n_*\}$ moments are determined from $v_{\sigma_*}(0) \neq 0$ (which can be normalized arbitrarily) . The higher order moments $\{v_{\sigma_*}(\rho)| \rho \geq n_*+1\}$ are determined by the exact (OPPQ) identities
\begin{eqnarray}
 \int dx\  x^{\sigma_*}{\cal O}_\sigma^{(n_*+q)}(x^2) \Phi_\sigma(x) = 0,\  for \ q \geq 1, \nonumber \\
 \sum_{i = 0}^{n_*+q} \Xi_{\sigma_*;i}^{(n_*+q)} v_{\sigma_*}(i)= 0 \nonumber \\
 v_{\sigma_*}(n_*+q) = -{1\over {\Xi_{\sigma;n_*+q}^{(n_*+q)}}} \sum_{i = 0}^{n_*+q-1} \Xi_{\sigma_*;i}^{(n_*+q)} v_{\sigma_*}(i).
\end{eqnarray}

\bigskip\noindent\underbar{QES states: Approach- II}
\smallskip

An alternative approach is to not determine the QES states from Eq.(41) but actually use the moment equation in Eq.(38) to generate linear constraints amongst the $\{v_\sigma(\rho)| 0 \leq \rho \leq n_*\}$ (i.e. they all depend on $v_\sigma(0)$), and amongst the $\{v_\sigma(\rho)| \rho \geq n_*+2\}$ with respect to the $\{v_\sigma(0), v_\sigma(n_*+1)$ moments (i.e. the linear dependence will be derived below). This is represented in Eq.(50). We now focus on deriving its form and applying OPPQ to it.

\subsection{QES Potential and Arbitrary (QES or non-QES) States: Generating the $v_{\sigma_*}(\rho)$ moments for $\rho \geq n_*+2$}

We now consider the moment equation for the QES-potential case and for all states of the QES  symmetry class  $\sigma = \sigma_*$. Our primary motivation is to show that OPPQ-$\Phi$ will recover the exact QES energies, for all $N \geq  d+1$,  with respects to the $\{v_{\sigma_*}(\rho)|\rho \geq n_*+1\}$ moments. This corresponds to an $m_s = 1$ problem. However,  in the $N\rightarrow \infty$ limit the OPPQ determinant (of the underlying $2 \times 2$ matrix) also generates other energy roots not related to the QES states. These will correspond to the non-QES states, and  exponentially converge to the true energies in the infinite limit.  In this approach, we are ignoring that Eq.(43) also tells us that the QES states are the roots of the BD polynomials. 

 As previously noted, the moment equation under the QES-potential condition in Eq.(41), and for the $\sigma_*$ parity states (QES or non-QES)  does not have a uniform $m_s$ index. The  first $n_*+1$ moments are linearly connected to $v_{\sigma_*}(0)$, thus defining an effective $m_s = 0$ relationship; whereas all the other moment are linearly related to $\{v_{\sigma_*}(n_*+1), v_{\sigma_*}(n_*)\}$, or equivalently  $\{v_{\sigma_*}(n_*+1), v_{\sigma_*}(0)\}$; thereby defining an effective $m_s = 1$ problem.  We are explicitly not using the fact that for the non-QES states: ($v_{\sigma_*}(0) = 0$.  For simplicity, we further abbreviate the notation for the relevant coefficient functions:

\begin{eqnarray}
{ C}_1(\rho+1) = m -{{b^2}\over {4g}} + \sqrt{g}(4\rho + 3 + \sigma_*) \nonumber \\
{ C}_0(\rho) = E- {b\over {\sqrt{g}}}(2\rho+{1\over 2} + \sigma_*), \nonumber \\
 { C}_{-1}(\rho-1) = 2\rho(2\rho-1+2\sigma_*).\nonumber\\
 \end{eqnarray}
The moment equation for the QES-symmetry class states (QES and non-QES states) then becomes:
\begin{eqnarray}
v_{\sigma_*}(\rho+1) =  {{C_0(\rho)}\over{C_1(\rho+1)}}v_{\sigma_*}(\rho)+{{C_{-1}(\rho-1)}\over{C_1(\rho+1)}} v_{\sigma_*}(\rho-1),\nonumber\\
\end{eqnarray}
for $0 \leq \rho \leq n_*-1$ and $\rho \geq n_*+1$, separately.  The recursive nature of Eq.(55) for $0 \leq \rho \leq n_*-1$ defines the relation 

\begin{equation}
v_{\sigma_*}(\rho) = M_{E,\sigma_*}(\rho,0) v_{\sigma_*}(0),   0 \leq \rho \leq n_* .
\end{equation}

From Eq.(55) we see that $v_{\sigma_*}(n_*+2)$ is generated through the linear superposition of$\{v_{\sigma_*}(n_*+1), v_{\sigma_*}(n_*)\}$. In general, we can express all the $\{v_{\sigma_*}(\rho)|\rho \geq n_*\}$ moments in terms of the linear sum of $\{v_{\sigma_*}(n_*), v_{\sigma_*}(n_*+1)\}$:

\begin{eqnarray}
v_{\sigma^*}(\rho) =  M_{E,\sigma_*}(\rho,n_*) v_{\sigma^*}(n_*) + M_{E,\sigma_*}(\rho,n_*+1) v_{\sigma^*}(n_*+1), \nonumber   \\
\end{eqnarray}
for $\rho \geq n_*$, where 
\begin{eqnarray}
M_{E,\sigma_*}(n_*,n_*) =  1 \nonumber \\
M_{E,\sigma_*}(n_*,n_*+1) = 0 \nonumber \\
M_{E,\sigma_*}(n_*+1,n_*) = 0 \nonumber \\
M_{E,\sigma_*}(n_*+1,n_*+1) = 1 . \nonumber \\
\end{eqnarray}
Inserting Eq.(57) into Eq.(55),  and making use of the independence of $\{v_{\sigma^*}(n), v_{\sigma^*}(n+1)\}$, gives:

\begin{eqnarray}
M_{E,\sigma_*}(\rho+1,\ell) =  {{C_0(\rho)}\over{C_1(\rho+1)}}M_{E,\sigma_*}(\rho,\ell)+{{C_{-1}(\rho-1)}\over{C_1(\rho+1)}} M_{E,\sigma_*}(\rho-1,\ell),\nonumber\\
\end{eqnarray}
for $\rho \geq n_*+1$ and $\ell = n_*,n_*+1$ subject to the initialization conditions in Eq.(58). Thus $M_{E,\sigma_*}(n_*+2,n_*) =  {{C_{-1}(n_*)}\over{C_1(n_*+2)}}$ and
$M_{E,\sigma_*}(n_*+2,n_*+1) =  {{C_0(n_*+1)}\over{C_1(n_*+2)}}$, yielding $v_{\sigma^*}(n_*+2) =  {{C_{-1}(n_*)}\over{C_1(n_*+2)}} v_{\sigma^*}(n_*) + {{C_0(n_*+1)}\over{C_1(n_*+2)}}v_{\sigma^*}(n_*+1)$.

Since $v_{\sigma_*}(n_*) = M_{E,\sigma_*}(n_*,0) v_{\sigma_*}(0)$, we have that 
\begin{eqnarray}
v_{\sigma_*}(\rho) & = & M_{E,\sigma_*}(\rho,n_*)M_{E,\sigma_*}(n_*,0) v_{\sigma^*}(0) + M_{E,\sigma_*}(\rho,n_*+1) v_{\sigma_*}(n_*+1), \nonumber   \\
\end{eqnarray}
or 
\begin{eqnarray}
v_{\sigma_*}(\rho) & = & M_{E,\sigma_*}(\rho,0) v_{\sigma^*}(0) + M_{E,\sigma_*}(\rho,n_*+1) v_{\sigma_*}(n_*+1), \rho \geq n_* \nonumber \\
\end{eqnarray}
where  $M_{E,\sigma_*}(\rho,0) = M_{E,\sigma_*}(\rho,n_*)M_{E,\sigma_*}(n_*,0)$. Also, it is implicitly understood that $M_{E,\sigma_*}(\rho,n_*+1) = 0$ for $0 \leq \rho \leq  n_*$.

Having defined Eq.(61), which effectively defines an $m_s = 1$ moment recursion relation, we want to implement Eq.(51), the OPPQ condition.

 Let ${\cal M}_{E,\sigma_*}(N,\ell) = \sum_{i=0}^N \Xi^{(N)}_{\sigma_*,i}M_{E,\sigma_*}(i,\ell)$ for $\ell = 0,n_*+1$. The OPPQ determinant condition becomes
\begin{eqnarray}
D_N(E) = Det\pmatrix{ {\cal M}_{E,\sigma_*}(N,0) & {\cal M}_{E,\sigma_*}(N,n_*+1)  \cr {\cal M}_{E,\sigma_*}(N+1,0) & {\cal M}_{E,\sigma_*}(N+1,n_*+1) } = 0, \nonumber \\
\end{eqnarray} 
for $ N\geq n_*+1$.

We know that the QES energies given by Eq.(43) must also satisfy Eq.(62), since it embodies the exact OPPQ conditions for these states. Therefore, the OPPQ determinant must factorize according to
\begin{eqnarray}
D_N(E) = P^{(n_*+1)}_{\sigma_*}(E) \times Poly_{N,\sigma_*}^{(Non-QES)}(E),
\end{eqnarray}

for $N \geq n_*+1$. That is, the first polynomial factor is that for the QES states in Eq.(43). The second polynomial factor's roots become the OPPQ converging approximants to the non-QES states. 
The numerical confirmation of this is given in Tables  9 and 10 where we compare the (exact) QES and non-QES energies generated through the above OPPQ analysis with the QES energies generated from the BD energy polynomial.

\begin{table}
\caption{\label{table1}
Comparison of QES and non-QES (of $\sigma_*$ symmetry) states computed through exact root formula $P^{n_*+1}_{\sigma_*}(E^*) = 0$ in Eq. (43) and
OPPQ-$\Phi$ Applied to OPPQ-(polynomial) determinant in Eq.(62-63). No rounding off for QES energies.
}
\centerline{
\begin{tabular}{lcccccc}
$N$  & $E_0^*$ & $E_2^*$ & $E_4^*$ & $E_6^*$  & $E_8$    & $E_{10}$ \\
     & -4.701631 & 2.289850 & 13.186912 & 28.822848 & NA & NA \\
\hline
\multicolumn{7}{c}{\vrule height14pt depth5pt width0pt $V(x) = x^6 + \sqrt{8}x^4-13 x^2$,  $n_*=3,\sigma_*=0$} \\
\hline
 4 & -4.701631 & 2.289850  &  13.186912  &  28.822848 &   &  \\
 5 & -4.701631 &  2.289850  & 13.186912  &  28.822848 &  49.879720 & \\
 6 & -4.701631 & 2.289850  & 13.186912 &  28.822848 &     47.994447 & 76.381590\\
 7 & -4.701631 &  2.289850 &  13.186912  &28.822848 &      47.679059 & 70.953850 \\
 8 & -4.701631& 2.289850  & 13.186912& 28.822848   &  47.624584 & 69.527914\\
 9 & -4.701631 & 2.289850  & 13.186912  &  28.822848 & 47.615172& 69.156251\\
 10 & -4.701631 & 2.289850 &13.186912 & 28.822848&      47.613408  &  69.058368\\
 11& -4.701631 &2.289850  & 13.186912 & 28.822848&   47.612358 &68.924938\\
\hline
\end{tabular}
}
\end{table}

\begin{table}
\caption{\label{table1}
Comparison of QES and non-QES (of $\sigma_*$ symmetry) states computed through exact root formula $P^{n_*+1}_{\sigma_*}(E^*) = 0$ in Eq. (43) and
OPPQ-$\Phi$ Applied to OPPQ-(polynomial) determinant in Eq.(62-63). No rounding off of QES energies.
}
\centerline{
\begin{tabular}{lcccccc}
$N $  & $E_1^*$ & $E_3^*$ & $E_5^*$ & $E_7^*$  & $E_9$    & $E_{11}$ \\
                                     & -6.629227& 4.618850 & 18.024593 & 34.897472 & NA & NA \\
\hline
\multicolumn{7}{c}{\vrule height14pt depth5pt width0pt $V(x) = x^6 + \sqrt{8}x^4-15 x^2$, $n_*=3,\sigma_*=1$} \\
\hline
 4 & -6.629227 & 4.618850  &  18.024593  &  34.897472 &   &  \\
 5 & -6.629227 & 4.618850 & 18.024593  &  34.897472&  56.9465755  & \\
 6 &-6.629227 & 4.618850  & 18.024593  &  34.897472 &     55.0485775& 84.3945915\\
 7 & -6.629227 &  4.618850 &  18.024593   &34.897472 &      54.7454855 &78.8505925 \\
 8 & -6.629227& 4.618850  &18.024593   & 34.897472  &  54.6960975 & 77.4351185\\
 9 & -6.629227 & 4.618850  & 18.024593   &  34.897472 & 54.6881985& 77.0885815\\
 10 & -6.629227 & 4.618850 &18.024593  & 34.897472&     54.6872425  &  77.0349865\\
 11& -6.629227 & 4.618850  & 18.024593  & 34.897472&    54.6873885 &77.0356645\\
\hline
\end{tabular}
}
\end{table}

\section{The Configuration Space QES Analysis}

We want to contrast the previous moment QES formulation with the conventional configuration space analysis. Although the configuration space analysis is easier to implement, its major deficiency is that it does not immediately transfer to the non-QES states. That is, the Bender-Dunne factorization property for their polynomials does not give any immediate information about the non-QES states, in contrast to the OPPQ factorization property expressed in Eq.(63).  This is primarily due to the inherent instability of the configuration space Hill determinant approach, which tries to quantize by imposing a truncation strategy to the ratio
 ${\Psi\over{\cal A}}= {{\Phi}\over{{{\cal A}^2}}} = \sum_{j=0}^\infty a_jx^j$. Although the moment's and configuration space representation generate the same QES- polynomials, the moment's formulation naturally truncates the  polynomials of degree greater than $n_*+1$ in Eq.(42) when defined in terms of the $v_{\sigma_*}(n)$'s; however, if the recursion relation in Eq.(46) is used, there is the misleading appearance that they can be defined up to degree $n_*+2$ based on the discussion pertaining to Eq.(47) (although $\tilde{\gamma}_{n_*+1} = 0$). All this is because the moment equation decouples the $v_{\sigma_*}(n_*+1)$ from the lower order moments; while, all the higher order moments (i.e. $v_{\sigma_*}(\rho)$, $\rho \geq n_{\sigma_*}+1$) couple to $v_{\sigma_*}(0)$ and $v_{\sigma_*}(n_*+1)$. This is not the case for the configuration space generated energy polynomials. One can generate them to all orders, as given by the power series expansion $a_j$'s.  The order of the recursion relation for the $a_j$'s stays the same (i.e. order {\it{one}}) regardless of the QES or non-QES nature of the state.  

Define the analytic function $P(x) = {\Psi\over{\cal A}}= {{\Phi}\over{{{\cal A}^2}}} = x^{\sigma_*}\sum_{i=0} c_i(E) x^{2i}$. The associated differential equation is:
\begin{eqnarray}
-\partial_x^2P(x) +\big({b\over {\sqrt{g}}} x + 2 \sqrt{g} x^3\big) \partial_x P(x) \nonumber  \\
+\  \Big((m+3\sqrt{g} -{{b^2}\over{4g}})x^2 - (E- {b\over{2\sqrt{g}}})\Big) P(x) = 0,
\end{eqnarray}
resuting in:

\begin{eqnarray}
(\sigma+2)(\sigma+1)c_1 =  \big( {b\over {\sqrt{g}}}({1\over 2}+\sigma) - E\big)  c_0, \nonumber \\
2(i+1)(2i+1+2\sigma) c_{i+1}  =   \big( {b\over {\sqrt{g}}}(2i +  {1\over 2}+\sigma) - E\big) c_i  \nonumber \\
+\big(  m -{{b^2}\over{4g}} +\sqrt{g}(4(i-1)+3+2\sigma)\big)c_{i-1} ,  i \geq 1.\nonumber \\
\end{eqnarray}

The coefficients are polynomials in the energy. For the power series to naturally truncate we want $c_I(E) = 0$ and the coeffieicnt of $c_{I-1}$ to be zero. This will make $c_{i+1} = 0$ for all $i \geq I$. If we call $I = n_*+1$, we recover the QES condition on the parameter and $c_{n_*+1}(E)$ becomes proportional to $P_{\sigma_*}^{(n_*+1)}(E)$.  We note that under the QES condition, since the coefficient of $c_{i-1}$ is zero, for $i = n_*+1$, the QES states correspond to $c_{ n_*+1}(E) = 0$.  However, these will always be the zeroes for the higher order polynomials, $c_{i+1}(E)$, for $i \geq n_*+1$. More importantly, if the potential function parameters satisfy the QES conditions, all the $\{c_i(E)|i \geq 0\}$ polynomials can be generated through a recursive, first order, relation. This is not the case for the $v_{\sigma_*}(\rho)$ energy-polynomials, since they naturally truncate at $\rho = n_*$ . Furthermore, the finite order recursion relation for these moments is not of uniform order, as argued in the previous sections.

If we move the $c_i$ term in Eq.(65) to the left hand side, we note that the recursive structure   is the reverse of the moment equation in Eq.(38), in the sense defined below. 

\begin{eqnarray}
\pmatrix{  m-{{b^2}\over{4g}} +\sqrt{g}(4\rho+3 + 2\sigma) \cr E-{b\over{\sqrt{g}}}(2\rho+{1\over 2} +\sigma) \cr 2\rho(2\rho-1+2\sigma)} \rightarrow \pmatrix{\rho  \rightarrow & i-1 \cr \rho   \rightarrow & i \cr \rho  \rightarrow & i+1} \rightarrow Coeff \pmatrix{c_{i-1}\cr c_i \cr c_{i+1}}. \nonumber\\
\end{eqnarray}
\vfil\break
That is, the recursive structure of the $c_i$'s, for $0\leq i \leq n_*+1$, produces the polynomial $c_{n_*+1}(E)$, which is the same as that generated by the $v_{\sigma_*}(\rho)$, for $0 \leq \rho \leq n_*$, and combined to produce the $P_{\sigma_*}^{(n_*+1)}(E)$ polynomial in Eqs.(42-43).

\section{The Bender-Dunne Sextic Potential}

We now consider the original Bender-Dunne Hamiltonian (with potential $V_{BD}$)
\begin{eqnarray}
H = -\partial_x^2 + {b\over { x^2}}  + m x^2 + x^6, \nonumber \\
b = {1\over 4} (4s-1)(4s-3) , \nonumber \\
m = -(4s+4J-2). \\
\end{eqnarray}
The wavefunction must assume the form $\Psi(x) = x^{\gamma} A(x^2)$, near the origin. Since the probability density must be integrable it follows that  $\gamma > -{1\over 2}$.  The indicial equation gives $\gamma^2-\gamma-b = 0$, or $\gamma = {{1\pm(4s-2)}\over 2}$. We take $\gamma = 2s-{1\over2}$.  Note that $A(x^2)$ suggests an analytic function of $x^2$ whreas ${\cal A}(x)$, as given below, corresponds to the leading asymptotic exponential form of the solution.

The QES states should assume the form: $\Psi(x) = x^\gamma P_d(x^2) {\cal A}(x)$, where the physical asymptotic factor is  ${\cal A}(x) = e^{-{{x^4}\over 4}}$. There  are only two ways to confirm this, algebraically. One is to implement the Hill representation truncation analysis to determine if such solutions exist. The other is to establish the existence of a $\Phi(x) = {\cal A}(x) \Psi(x)$ representation whose $\nu$-moment equation confirms the existence of such solutions, as was done for the sextic anharmonic oscillator potential  in the previous sections.  Within the $\Psi$ representation, an asymptotic analysis can suggest the potential function parameter constraints consistent with a QES type of solution.
Tailoring the  asymptotic analysis in Eq.(25) to the explict form of the BD potential yields $\Psi(x) \sim x^\delta  exp(-{1\over 4} x^4)$, where $\delta  = -{{(m+3)}\over 2}$. Here $\delta = \gamma + 2d$, since $P_d(x^2)$ is a polynomial of degree $2d$. That is,` 

A Hill representation truncation analysis for ${{\Psi(x)}\over {x^\gamma {\cal A}(x)}} \equiv C(x^2) = \sum_{i=0}^\infty c_i(E) x^{2i}$ gives

\begin{eqnarray}
c_0 = 1 \nonumber \\
c_1(E) = -{{E}\over {4\gamma+2}}  c_0 \nonumber\\
 c_{i+1}(E) = {{-e c_i(E) + (2\gamma+m+ 4i-1)c_{i-1}(E)}\over{(i+1)(4\gamma+4i +2)}} , i \geq 1. \nonumber \\
\end{eqnarray}
We see that if 
\begin{eqnarray}
4 n_* + 2\gamma+m+ 3  = 0,\ or  \ J = n_*+1,\nonumber \\
 c_{n_*+1}(E) = 0, 
\end{eqnarray}
 determines the QES states; and  $C(x^2) = Polynomial \ of \ degree \ x^{2n_*}\equiv P_{n_*}(x^2)$. That is $d = n_*$, or $\gamma+2n_* = -{{(m+3)}\over 2}$, consistent with the $\Psi$-asymptotic analysis above. 

As in the OPPQ-$\Psi$ analysis, we could develop a moment equation for $\Psi$, retaining the indicial exponent.  However, one's first inclination is to strip the indicial factor, in order to generate a less complicated analysis. We will do so for illustrative purposes, only. As we shall see, stripping the indicial factor is incorrect:  the {\it{Bessis}} representation is obtained by not only keeping the indicial factor but further enhancing it by an additional indicial factor: $\Phi(x) = \Psi(x)x^\gamma {\cal A}(x)$, or $\Phi(x) = P_d(x^2)x^{2\gamma} {\cal A}^2(x)$.  We note that the latter is multiplying $\Psi(x)$ by its leading asymptotic form as $x \rightarrow \infty$ as well as $x \rightarrow 0$.

 Before examining the Bessis representation, we implement OPPQ on two representations. The first of these involves stripping the wavefunction of the indicial factor: $A(x^2) = x^{-\gamma}\Psi(x)$. The second will be to enhance this by multiplying by the physical (exponentially decaying) asymptotic form, ${\tilde\Phi}(x^2) = x^{-\gamma}\Psi(x) {\cal A}(x^2)$.

For the first case, we work with the even power moments of $A(x^2)$: $u(\rho) \equiv \int_0^\infty dx x^{2\rho} A(x^2)$. The relevant differential equation is

\begin{eqnarray}
-\partial_x^2A - {{2\gamma}\over x} \partial_x A + (m x^2 +x^6)A = E A.
\end{eqnarray}
Upon multiplying both sides by $x^{2\rho+2}$ and integrating by parts we obtain the moment equation

\begin{eqnarray}
u(\rho+4) = - mu(\rho+2) + E u(\rho+1)+ 2(\rho+1-\gamma)(2\rho+1) u(\rho)  , \rho \geq 0. \nonumber \\
\end{eqnarray}
The missing moment structure $m_s = 3$, resulting in $u(\rho) = \sum_{\ell = 0}^3M_E(\rho,\ell) \ u(\ell)$. The OPPQ analysis is done with respects to the representation $A(x^2) = \sum_{j=0}^\infty \Omega_j {\cal P}^{(j)}(x) {\cal A}(x)$. The data in Table 11  gives the results for $n_* = 3, J = n_*+1 = 4, s = 1$. We emphasize that our objective is not to show the full convergence of the non-QES states, which becomes manifest at higher orders (i.e. $N\rightarrow \infty$), but to suggest the veracity of our OPPQ analysis as applied to both QES and non-QES states.
\vfil\break
\begin{table}
\caption{\label{table1}
Comparison of QES and non-QES states computed through exact root formula $c_4(E) = 0$ in Eq. (70) and
OPPQ-($x^{-\gamma}\Psi$) for $m_s = 3$ moment equation in Eq.(72). Parameters $s = 1$, $J = n_*+1$, and $n_* = 3$.
}
\centerline{
\begin{tabular}{lcccccc}
$N $  & $E_0^*$ & $E_1^*$ & $E_2^*$ & $E_3^*$  & $E_4$    & $E_{5}$ \\
                                     & -20.926277 & -6.487752 & +6.487752 & +20.926277& NA & NA \\
\hline
\multicolumn{7}{c}{\vrule height14pt depth5pt width0pt $V(x) = x^6 + mx^2 + {b\over{x^2}}$, $b = 3/2$, m = -18} \\
\hline
1 & -17.752051&  & & & &\\
2 &-23.465769 & -5.699531& & & & \\
3 & -20.857859& -8.880996& 4.319160& & &  \\
4 & -20.926277& -6.487752 & 6.487752 & 20.926277 & &\\
 5& -20.926277& -6.487752 & 6.487752  & 20.926277 & 52.309013  &  \\
 6& -20.926277& -6.487752 & 6.487752  & 20.926277 & 41.490341 & 94.456407 \\
 7& -20.926277& -6.487752 & 6.487752  & 20.926277 &      38.426546 & 71.311307\\
8& -20.926277& -6.487752 & 6.487752  & 20.926277 &      37.787371 &61.916009 \\
9& -20.926277& -6.487752 & 6.487752  & 20.926277 &  37.839537 & 58.167011\\
36&                 &                   &                   &                   &  38.002392718 & 57.536940282                   \\
\hline
\end{tabular}
}
\end{table}

 The second OPPQ analysis is done on  ${\tilde\Phi}(x^2) = x^{-\gamma}\Psi(x) exp(-{{x^4}\over 4})$, which involves the previous representation multiplied by an additional exponential asymptotic form.  
We obtain the differential equation for ${\tilde\Phi}(x^2)$ 
\begin{eqnarray}
x \partial_x^2{\tilde\Phi} +2 \big( \gamma +x^4\big)\partial_x {\tilde\Phi} +(2\gamma-m+3\big) x^3{\tilde\Phi} +E x {\tilde\Phi} = 0. \nonumber \\
\end{eqnarray}
Upon multiplying both sides by $x^{2\rho+1}$, and defining $u(\rho) = \int dx \ x^{2\rho}{\tilde\Phi}(x^2)$, we obtain the moment equation:

\begin{eqnarray}
\big( 4\rho -2\gamma+m+7\big) u(\rho+2) = E u(\rho+1) +  2(2\rho+1)( \rho+1-\gamma) u(\rho), \rho \geq 0. \nonumber \\
\end{eqnarray}
So long as $\gamma \neq integer$, we can generate all the power moments and pursue OPPQ for generating the exact QES and (converging) approximate non-QES.
The OPPQ representation in this case is ${\tilde \Phi}(x) =\sum_{j=0}^\infty \Omega_j {\cal O}^{(j)}(x) {\cal A}^2(x)$, where ${\cal O}^{(j)}(x)$ are the orthonormal polynomials of ${\cal A}^2(x)$.
The results are given in Table 12. The convergence of the non-QES is much faster. The above moment equation almost suggests the manifest existence of QES solutions.
However it is not a three term recursion relation, since the effective missing moment order is $m_s = 1$.

A third OPPQ analysis (the Bessis representation) is possible on a somewhat different moment equation formulation. Consider $\Phi(x) = \Psi(x) x^\gamma exp(-{{x^4}\over 4})$. This is no longer an analytic function at the origin: $\Phi(x) \approx O( x^{2\gamma}) $, recall $\gamma > -{1\over 2}$.  The differential equation is that of Eq.(73) with $\gamma \rightarrow -\gamma$, plus an additional term (due to a variant on the indicial equation) yielding:

\begin{eqnarray}
x \Phi''(x) +2 \big( -\gamma +x^4\big) \Phi'(x) +(-2\gamma-m+3\big) x^3\Phi(x) +E x \Phi(x) + 2{\gamma \over x}\Phi(x) = 0. \nonumber \\
\end{eqnarray}
If we multiply by $x^{2\rho+1}$ and integrate over the nonnegative real axis, (i.e. $\int_\epsilon^\infty dx$ , $\epsilon \rightarrow 0^+$)  we obtain

\begin{eqnarray}
\int_\epsilon^\infty dx \ x^{2\rho+1}\Big(Eq.(75))\Big) = 
&-\Big( \big(2 \epsilon^{2\rho+5} -2(\gamma+\rho+1)\epsilon^{2\rho+1}\big )\Phi(\epsilon) + \epsilon^{2\rho+2} \Phi'(\epsilon)\Big) \nonumber \\
&+ \int_{\epsilon}^\infty dx \Big( (\rho+1) \big( 4\rho+2+4\gamma\big) x^{2\rho} + E x^{2\rho+2} \Big)\Phi(x) \nonumber \\
&-\int_{\epsilon}^\infty dx \Big(4\rho+2\gamma+m +7\Big) x^{2\rho+4} \Phi(x) .\nonumber \\
\end{eqnarray}
Since $\Phi(\epsilon) = \epsilon^{2\gamma}(1+O(\epsilon^2))$, $\Phi'(\epsilon)= 2\gamma\epsilon^{2\gamma-1} +2(\gamma+1)O(\epsilon^{2\gamma+1})$, the first term in Eq.(76) vanishes in the zero limit : $\lim_{\epsilon\rightarrow 0}\Big( \big(2 \epsilon^{2\rho+5} -2(\gamma+\rho+1)\epsilon^{2\rho+1}\big )\Phi(\epsilon) + \epsilon^{2\rho+2} \Phi'(\epsilon)\Big) = 0$, for $\rho \geq -1$ and $\gamma > -{1\over 2}$. Additionally, the integral expressions are finite for $\rho \geq -1$. We therefore we have the following moment equation, valid for $\gamma > -{1\over 2}$ and $\rho \geq -1$,  where $\nu(p) \equiv \int_0^\infty dx \ x^p \Phi$:
\begin{eqnarray}
\big( 4\rho +2\gamma+m+7\big) \nu(\rho+2) = E\nu(\rho+1) +  (\rho+1) \big( 4\rho+2+4\gamma\big) \nu(\rho), \rho \geq -1. \nonumber \\
\end{eqnarray}
or $(\rho \rightarrow \rho+1)$:

\begin{eqnarray}
\big( 4\rho +2\gamma+m+3\big) \nu(\rho+1) = E \nu(\rho) +  \rho \big( 4\rho-2+4\gamma\big) \nu(\rho-1), \rho \geq 0. \nonumber \\
\end{eqnarray}
This is also a three term recursion relation in which the QES potential function conditions are manifest. That is, if $\gamma+{{m+3}\over 2} = -2n_*$, where $\gamma = 2s-{1\over 2}$, then only the 
first $n_*+1$ moments can be generated $\{\nu(\rho) | 0 \leq \rho \leq n_*\}$, all defining an effective $m_s = 0$ missing moment problem in which the corresponding moments become polynomials in the energy (i.e. $\nu(0) = 1$), $\nu(\rho) = Polynomial \ of \ degree\  \rho  \ in \ E$. The non-QES states must have these first $n_*+1$ moments identically zero:

\begin{eqnarray}
\nu_{QES}(\rho) \neq 0 , 0 \leq \rho \leq n_*, \nonumber \\
\nu_{non-QES}(\rho) = 0, 0 \leq \rho \leq n_*, \ if \ V_{BD}\ admits\ QES\ states.
\end{eqnarray}

As in the sextic anharmonic oscillator case, the $\nu(n_*+1)$ moment decouples from the moment equation.  We can repeat all the different types of OPPQ computational implementations done for the sextic anharmonic oscillator; however, we are only interested in repeating the OPPQ computational analysis that uniformly generates the QES and the non-QES states. 

As in the sextic anharmonic oscillator case, if the $V_{BD}$ potential admits QES states, then the $\{\nu(\rho)|\rho\geq n_*+2\}$ moments couple to the $\{\nu(n_*+1),\nu(n_*)\}$ moments, through an effective $m_s = 1$ recursion relation. However, $\nu(n_*)$ couples to all the lower order moments through an $m_s = 0$ recursion relation. Therefore, the $\{\nu(\rho)|\rho \geq n_*+2\}$ effectively couple, through an $m_s = 1$ relation, to $\{\nu(n_*+1),\nu(0)\}$ . We can apply OPPQ on this relation and uniformly obtain the QES and non-QES states.  That is, we are not using the BD energy polynomials,these are contained within the OPPQ conditions. The following discussion defines  the necessary relation connecting the $\{\nu(\rho)|\rho \geq n_*+2\}$ moments  to the $\{\nu(n_*+1),\nu(0)\}$ moments.

Within the Bessis representation $\Phi(x) = \Psi(x) x^\gamma {\cal A}(x)$, ${\cal A}(x) = e^{-{x^4}\over 4}$, the OPPQ representation becomes
$\Phi(x) =  \sum_{j=0}^\infty \Omega_j {\cal Q}^{(j)}(x) x^{2\gamma}{\cal A}^2(x)$, where the ${\cal Q}^{(j)}(x)$ are the orthonormal polynomials of 
${\cal B}(x) \equiv x^{2\gamma}{\cal A}^2(x)$.  We will work on the half real axis in terms of the $x^2$ variable. We make explicit this $x^2$ dependence, $\Phi(x) \rightarrow\Phi(x^2)$, ${\cal A}(x) \rightarrow {\cal A}(x^2)$, ${\cal B}(x) \rightarrow {\cal B}(x^2)$, and ${\cal Q}^{(j)}(x) \rightarrow {\cal Q}^{(j)}(x^2)$.

Define  ${\cal A}_{\sigma}(x^2) = exp(-{{x^4}\over {\sigma}})$, where $\sigma = 2$. The orthonormality property for the  ${\cal Q}^{(j)}(x^2)$'s becomes
$\int_0^\infty dx {\cal Q}^{(j_1)}(x^2) {\cal Q}^{(j_2)}(x^2) {\cal B}(x^2) =  \int_0^\infty \ d\xi {\cal Q}^{(j_1)}(\xi) {\cal Q}^{(j_2)}(\xi) {{{\cal B}(\xi) }\over {2\sqrt{\xi}}} = \delta_{j_1,j_2}$, 
where $\xi = x^2$. These orthonormal polynomials are generated from the moments  $m(\rho) = \int_0^\infty d\xi \xi^\rho {{{\cal B}(\xi) }\over {2\sqrt{\xi}}}$. The weight becomes
${{{\cal B}(\xi) }\over {2\sqrt{\xi}}} = {{\xi^{\gamma-{1\over 2}}}\over 2} \exp(-{{\xi^2}\over {\sigma}})$. 
Recalling that $\gamma = 2s-{1\over 2}$,  we obtain   $m(\rho) = {1\over 2}\int d\xi \xi^{\rho+2s-1} \exp(-{{\xi^2}\over {\sigma}}) = {1\over 4} \int d\zeta  \zeta^{{\rho\over 2}+s-1} exp(-{{\zeta}\over {\sigma}}) ={1\over 4} 2^{{\rho\over 2}+s}\Gamma ({\rho\over 2}+s)$, having set ${\sigma} = 2$. This enables us to generate the orthogonal polynomials, ${\cal Q}^{(j)}(\xi) = \sum_{i=0}^j\Xi_i^{(j)}\xi^i$. 

We now repeat the OPPQ analysis we did for the sextic anharmonic oscillator. The $\{\nu(\rho)|0\leq \rho \leq n_*\}$ moments satisfy an $m_s = 0$ moment equation regardless of the nature of the discrete state. This is true for the QES states. This is true for the non-QES states when the potential function satisfies the QES condition; although in this case, they are identically zero (in the following analysis we do not impose this, but it will be the result as the OPPQ quantization order goes to infinity, $N \rightarrow \infty$). If the potential function does not satisfy the QES conditions, then all the states satisfy an $m_s = 0$ moment equation, to all order. Accordingly, we have:

\begin{eqnarray}
\nu(\rho) =  M_E(\rho,0) \nu(0),\ 0 \leq \rho \leq n_*, \nonumber \\
M_E(0,0) \equiv 1.
\end{eqnarray}
All the moments of order $n_*+2$ or higher, are linearly dependent on the moments $\{\nu(n_*),\nu(n_*+1)\}$:
 
\begin{eqnarray} 
\nu(\rho) = \sum_{\ell = n_*}^{n_*+1} {\cal N}_E(\rho,\ell) \nu(\ell),\ n_*+2 \leq \rho < \infty, \nonumber \\
{\cal N}_E(\ell_1,\ell_2) = \delta_{\ell_1,\ell_2}, \ n_* \leq \ell_{1,2} \leq n_*+1,
\end{eqnarray}
where ${\cal N}_E(\rho,\ell)$ satisfies Eq.(78) for $n_*+2\leq \rho < \infty$. 

Finally, we combine these to produce the representation
\begin{eqnarray}
\nu(\rho) = \sum_{\ell = 0,n_*+1} M_E(\rho,\ell) \nu(\ell),\ 0 \leq \rho <\infty,
\end{eqnarray}
where
\begin{eqnarray}
M_E(\rho,0) = \ determined\ from\ m_s \ = 0 \ moment \ equation\ for\ 0 \leq \rho \leq n_*, \nonumber \\
M_E(\rho,n_*+1) \equiv 0, \ 0 \leq \rho \leq n_*, \nonumber \\
M_E(n_*+1,0)\equiv 0, M_E(n_*+1,n_*+1)\equiv 1, \nonumber \\
M_E(\rho,0) = {\cal N}_E(\rho,n_*) M_E(n_*,0),\rho \geq n_*+2, \nonumber\\
M_E(\rho,n_*+1) = {\cal N}_E(\rho,n_*+1), \ \rho \geq n_* +2. \nonumber \\
\end{eqnarray}

We  implement OPPQ by demanding that 

\begin{eqnarray}
\int d\xi {\cal Q}^{(N+\ell_r)}(\xi) \Phi(\xi) = 0, N\geq n_*+1, \ and \ \ell_r = 0,1; \nonumber \\
\sum_{i=0}^{N+\ell_r}\Xi_i^{(N+\ell_r)} \nu(i) = 0,\nonumber \\
\sum_{\ell_c =0,n_*+1}\Big(\sum_{i=0}^{N+\ell_r}\Xi_i^{(N+\ell_r)}M_E(i,\ell_c) \Big) u(\ell_c) = 0.
\end{eqnarray}
The latter results in a $2 \times 2$ set of simultaneous equations whose determinant exhibits the factorized form $D_N(E) = Poly_{QES}(E) \times Poly_{nonQES}(E)$. The QES polynomial factor contains all the QES roots consitent with the BD energy polynomial. The other polynomial factor generates the approximate non-QES energies through its roots that converge, exponentially fast, to the true non-QES values.    The results of this analysis are given in Table 13, with a much improved convergence compared to the case reflected in Table 12.

\section{ Conclusion}

We have presented an extensive OPPQ analysis of the QES and non-QES states for the sextic anharmonic oscillator and the Bender and Dunne sextic potential. The OPPQ analysis in either the $\Psi$ representation or $\Phi$ representation yields the exact QES states and approximates the non-QES states (through converging approximants). Within the Bessis function representation ($\Phi$) we can recover the configuration space Bender and Dunne energy orthogonal polynomials, leading to exact formulas for the energies, as well as the wavefunctions. We have shown that the reason for the singular behavior (breakdown) of the Bender and Dunne orthogonal polynomials is due to the breakdown of the order of the moment equation in the Bessis representation. This moments' intepretation was known by Handy and Bessis within the context of their formulation of the Eigenvalue Moment Method.  This breakdwon in the moment equation's order  can be interpreted as a spontaneous breakdown of the implicit degree of freedom within the moment's representation. The OPPQ moments' representation also reveals additional structure for the non-QES states (i.e. lower order moments are zero within the Bessis representation). We believe these propeties extend to multidimensional systems. Although we have not proved that all one dimensional QES systems must have an $m_s = 0$ moment equation for the QES states (within the Bessis representation), we believe that the two examples presented here strongly argue in favor of this.

\begin{table}
\caption{\label{table1}
Comparison of QES and non-QES states computed through exact root formula $c_4(E) = 0$ in Eq. (70) and
OPPQ-${\tilde{\Phi}}$ for $m_s = 1$ moment equation in Eq.(74).  Parameters $s = 1$, $J = n_*+1$, and $n_* = 3$.
}
\centerline{
\begin{tabular}{lcccccc}
$N $  & $E_0^*$ & $E_1^*$ & $E_2^*$ & $E_3^*$  & $E_4$    & $E_{5}$ \\
                                     & -20.926277 & -6.487752 & +6.487752 & +20.926277& NA & NA \\
\hline
\multicolumn{7}{c}{\vrule height14pt depth5pt width0pt $V(x) = x^6 + mx^2 + {b\over{x^2}}$, $b = 3/2$, m = -18} \\
\hline
1 &-12.552595 &  & & & &\\
2 &-19.663222 &  -9.597580& & & & \\
3 &  -20.883219 & -6.093770 & 9.002550 & & &  \\
4 & -20.926277& -6.487752 & 6.487752 & 20.926277 & &\\
 5& -20.926277& -6.487752 & 6.487752  & 20.926277 & 36.988059  &  \\
 6& -20.926277& -6.487752 & 6.487752  & 20.926277 & 37.544189 &  58.584676 \\
 7& -20.926277& -6.487752 & 6.487752  & 20.926277 &      37.887188   & 56.623863\\
8& -20.926277& -6.487752 & 6.487752  & 20.926277 &    37.976840  &57.031923 \\
9& -20.926277& -6.487752 & 6.487752  & 20.926277 &  37.996662 & 57.372312 \\
\hline
\end{tabular}
}
\end{table}

\begin{table}
\caption{\label{table1}
Comparison of QES and non-QES states computed through exact root formula $c_4(E) = 0$ in Eq. (70) and
OPPQ-$\Phi$ for $m_s = 1$ moment equation in Eq.(78).  Parameters $s = 1$, $J = n_*+1$, and $n_* = 3$.
}
\centerline{
\begin{tabular}{lcccccc}
$N $  & $E_0^*$ & $E_1^*$ & $E_2^*$ & $E_3^*$  & $E_4$    & $E_{5}$ \\
                                     & -20.926277 & -6.487752 & +6.487752 & +20.926277& NA & NA \\
\hline
\multicolumn{7}{c}{\vrule height14pt depth5pt width0pt $V(x) = x^6 + mx^2 + {b\over{x^2}}$, $b = 3/2$, m = -18} \\
\hline
4 & -20.926277& -6.487752 & 6.487752 & 20.926277 & &\\
 5& -20.926277& -6.487752 & 6.487752  & 20.926277 & 40.921277 &  \\
 6& -20.926277& -6.487752 & 6.487752  & 20.926277 & 38.584899&   67.221602  \\
 7& -20.926277& -6.487752 & 6.487752  & 20.926277 & 38.122298  & 60.484136 \\
8& -20.926277& -6.487752 & 6.487752  & 20.926277 &  38.026662   & 58.428088\\
9& -20.926277& -6.487752 & 6.487752  & 20.926277 &  38.007296 &  57.788594 \\
10& -20.926277& -6.487752 & 6.487752  & 20.926277 & 38.003397   & 57.603593\\
11& -20.926277& -6.487752 & 6.487752  & 20.926277 &  38.002606 &  57.554137 \\
\hline
\end{tabular}
}
\end{table}


\vfil\break

\section*{Acknowledgments}

Discussions with  Dr. D. Bessis are greatly appreciated. One of the authors (DV) is grateful for the support received from the National Science Foundation
through a grant for the Center for Research on Complex Networks (HRD-1137732).

\end{document}